\documentclass[fleqn,usenatbib]{mnras}

\usepackage{newtxtext,newtxmath}

\usepackage[T1]{fontenc}
\usepackage{ae,aecompl}
\usepackage{tabularx}
\usepackage{caption}

\usepackage{graphicx}	
\usepackage{subfig}

\title[Bar Formation \& Evolution in Auriga]{Bar formation and evolution in the cosmological context: Inputs from the Auriga simulations}

\author[F. Fragkoudi et al.]{Francesca Fragkoudi$^{1}$\thanks{E-mail:francesca.fragkoudi@durham.ac.uk}, Robert J. J. Grand$^{2}$, R\"{u}diger Pakmor$^{3}$,
Facundo G\'{o}mez$^{4,5}$ \and Federico Marinacci$^{6}$, Volker Springel$^{3}$ \\
\\
$^{1}$Institute for Computational Cosmology, Department of Physics, Durham University, South Road, Durham DH1 3LE, UK \\
$^{2}$Astrophysics Research Institute, Liverpool John Moores University, 146 Brownlow Hill, Liverpool, L3 5RF, UK \\
$^{3}$Max-Planck-Institut f{\"u}r Astrophysik, Karl-Schwarzschild-Str. 1, 85748 Garching, Germany \\
$^{4}$Instituto de Investigaci\'on Multidisciplinar en Ciencia y Tecnolog\'ia, Universidad de La Serena, Ra\'ul Bitr\'an 1305, La Serena, Chile\\
$^{5}$Departamento de Astronom\'ia, Universidad de La Serena, Av. Juan Cisternas 1200 Norte, La Serena, Chile \\ 
$^{6}$Department of Physics \& Astronomy ``Augusto Righi'', University of Bologna, via Gobetti 93/2, 40129 Bologna, Italy\\%
}

\date{Accepted -. Received -; in original form -}

\pubyear{2024}

\begin{document}

\label{firstpage}
\pagerange{\pageref{firstpage}--\pageref{lastpage}}
\maketitle

\begin{abstract}
Galactic bars drive the internal evolution of spiral galaxies, while their formation is tightly coupled to the properties of their host galaxy and dark matter halo. To explore what drives bar formation in the cosmological context and how these structures evolve throughout cosmic history, we use the Auriga suite of magneto-hydrodynamical cosmological zoom-in simulations. We find that bars are robust and long-lived structures, and we recover a decreasing bar fraction with increasing redshift which plateaus around $\sim20\%$ at $z\sim3$. We find that bars which form at low and intermediate redshifts grow longer with time, while bars that form at high redshifts are born `saturated' in length, likely due to their merger-induced formation pathway. This leads to a larger bar-to-disc size ratio at high redshifts as compared to the local Universe.
We subsequently examine the multi-dimensional parameter space thought to drive bar formation. We find that barred galaxies tend to have lower Toomre $Q$ values at the time of their formation, while we do not find a difference in the gas fraction of barred and unbarred populations when controlling for stellar mass. Barred galaxies tend to be more baryon-dominated at all redshifts and assemble their stellar mass earlier, while galaxies that are baryon-dominated but that do not host a bar, have a higher ex-situ bulge fraction. We explore the implications of the baryon-dominance of barred galaxies on the Tully-Fisher relation, finding an offset from the unbarred relation; confirming this in observations would serve as additional evidence for dark matter, as this behaviour is not readily explained in modified gravity scenarios.
\end{abstract}




\section{Introduction}

The question of why some galaxies form a stellar bar and others do not, has puzzled astronomers for over a century now.
Since the early studies on the morphology of `extra-galactic nebulae', \cite{Curtis1918} noted that some spiral galaxies have \emph{``a band of matter extending diametrically across the nucleus and inner parts of the spiral''}, which he dubbed the `$\phi$-type spirals', due to their rough similarity to the Greek letter $\phi$. Hubble later classified galaxies into the now commonly used `Hubble Sequence', separating spiral galaxies into what he called the `normal' spirals, and the `barred' spirals \citep{Hubble1926,Hubble1936}.
We now know that such barred spirals are in fact as common, if not more so, than unbarred spirals in the local Universe, with some studies finding bar fractions as high as 70\%, depending on the wavelength and method used to identify them \citep{Eskridgeetal2000,Menendezetal2007,Aguerrietal2009,Gadotti2009,Mastersetal2011,Erwin2018}.

While the fraction of barred galaxies is high in the local Universe, it is generally found to decrease towards higher redshifts (\citealt{Shethetal2008,Melvinetal2014,Simmonsetal2014}, but see also \citealt{Jogeeetal2004}). Studies using data from the Hubble Space Telescope find that  bars are present in about $\sim10\%$ of disc galaxies up to $z\sim2$ (e.g. \citealt{Simmonsetal2014}), while more recent studies using data from the James Webb Space Telescope (JWST) find multiple bars at $z>1$ (e.g. \citealt{Guoetal2023,Costantinetal2023}) and a higher bar fraction than previously reported: $\sim$20\% at $z\sim1-2$ and $\sim$15\% between $z\sim2-3$ \citep{LeConteetal2024}. Bars have also recently been identified out to redshifts of $z\sim4$ \citep{Smailetal2023,Tsukuietal2024,Amvrosiadisetal2024}. At the same time, studies exploring stellar populations in local galaxies find evidence of old bars in the local Universe (\citealt{Gadottietal2015,deSaFreitasetal2023}; de Sà-Freitas et al. in prep.), suggesting that these structures are robust and long-lived. The decreasing fraction of bars at higher redshifts, hints at that the tight link between bar formation and the evolution of spiral galaxies, while their presence in the early Universe suggests that spiral galaxies might be more dynamically mature at these early times than previously thought (see also e.g. \citealt{Rizzoetal2020,Lellietal2021}).

Bars affect their host galaxy in a variety of ways, and their formation marks a period in which internal evolutionary processes begin to dominate (e.g. \citealt{KormendyKennicutt2004}). As they evolve, bars are found to grow longer with time (e.g. \citealt{AthanassoulaMisiriotis2002,GhoshDiMatteo2024}), due to their ability to exchange angular momentum in the galaxy \citep{LyndenBellKalnajs1972}, i.e. by emitting angular momentum from material at resonance in the inner disc, which is then absorbed by resonant material in the outer disc and the spheroidal halo \citep{Sellwood1980,Athanassoula2002}.  
Bars affect the galaxy's evolution by funnelling gas within the bar to the central-most regions of the galaxy, where it can form nuclear structures such as nuclear discs and rings (e.g. \citealt{Athanassoula1992b, Knapenetal2002, Comeronetal2010, Ellisonetal2011, Fragkoudietal2016, Sormanietal2018, deLorenzoCaceresetal2019,MendezAbreuetal2019,Leamanetal2019,Gadottietal2019,Bittnereta2020}).
They also re-shape the inner regions of their host galaxies via the formation of a vertically extended bulge, referred to as an X-shaped or boxy/peanut (b/p) bulge, which forms due to vertical instabilities in the bar, is supported by resonant orbits \citep{Combesetal1990,Rahaetal1991,PfennigerFriedli1991,Patsisetal2002,MartinezValpuestaetal2006,Quillenetal2014,Portailetal2015,Fragkoudietal2015,SellwoodGerhard2020,LiShlosmanetal2023}, and which redistributes stars and gas in the inner regions (e.g. \citealt{Nessetal2013a,DiMatteoetal2015,Fragkoudietal2017b,Fragkoudietal2017c,Fragkoudietal2018,Debattistaetal2020}).

There are a number of properties which are known to be important in determining whether a bar can form or not, and which determine how it subsequently evolves. 
For example, an axisymmetric stellar disc will be unstable to local  instabilities if it is too `cold', i.e. if it does not have any \emph{random motions} \citep{Toomre1964}.
The early work of \cite{Hohl1971} explored the long-term stability of stellar discs using N-body simulations, to establish that even when a disc has enough velocity dispersion to prevent axisymmetric instabilities (i.e. when the Toomre stability criterion $Q>1$), it can still be unstable to global bar-forming modes. Subsequent work has shown that further increasing the velocity dispersion of a disc can delay bar formation, and lead to an overall weaker bar, but might not be enough to completely stabilise the disc against bar formation  (e.g. \citealt{AthanassoulaSellwood1986,Combesetal1990}).

Another fundamental parameter which sets whether a galaxy will form a bar, is its
\textit{disc-dominance}, i.e. how much the stellar disc dominates over the dark matter halo in the inner regions. In this work we will interchangeably use either `disc-dominance' or `baryon-dominance' to refer to this, as both are used in the literature. \cite{OstrikerPeebles1973} showed that embedding an otherwise unstable disc within a massive halo, can stabilise it against bar formation; the existence of a large number of unbarred galaxies in the Universe therefore provided an early indication of the existence of such massive halos surrounding spiral galaxies. The disc-dominance of a galaxy is usually defined as the disc's contribution to the circular velocity curve at a given characteristic radius (e.g. at 2.2 disc scale-lengths \citealt{Sackett1997}; see also \citealt{Efstathiouetal1982}) -- and as such depends on the enclosed mass within that radius.
Later works further expanded on the interaction between bars and halos, such as the seminal paper of \cite{Athanassoula2002}, which showed that while dark matter halos can stabilise the disc against bar formation for an initial period of time, the existence of a \emph{live} dark matter halo in fact helps bars grow longer and stronger later in their evolution. This occurs because a `live' halo is able to absorb the angular momentum emitted by the disc via the bar, whereas in the case of a disc embedded within a static halo, only the outer disc is available to act as a `absorber' of angular momentum. This `emission' and 'absorption' of angular momentum in the galaxy is driven by the bar, i.e. the inner regions of the disc emit angular momentum -- which leads to a longer and stronger bar -- while the outer disc and halo absorb the angular momentum, leading to a more extended disc and to an increase in net rotation in the halo. It is important to note that the velocity dispersion of the halo -- not just its mass -- play a role in the angular momentum transfer \citep{Athanassoula2003}, which further complicates matters.

While the dark matter halo is the most massive `spheroid' in the galaxy, a \emph{spheroidal bulge component} in the central regions will also contribute to stabilising the disc against bar formation \citep{OstrikerPeebles1973}, while also acting as an absorber of angular momentum from the bar (e.g. \citealt{Sahaetal2012}). The resonant processes driving this redistribution of angular momentum are similar to those taking place between the bar and the halo; the absorption of angular momentum by the bulge/spheroid reduces the angular momentum of the disc and can lead to the spin-up and flattening of an initially non-rotating spheroidal bulge (e.g. \citealt{Sahaetal2012,Sahaetal2016}).

Finally, the \emph{gas fraction} of galaxies has long been thought to play a role in the formation and subsequent evolution of bars. The works of \cite{BournaudCombes2002} and \cite{Bournaudetal2005}, which explored bar formation in N-body and hydrodynamics simulations, suggested that bars can be weakened or even completely destroyed in the presence of gas. They attributed this to two mechanisms; first, the formation of a central mass concentration (CMC) due to bar-induced gas inflows, which weakens the bar, and second, the transport of angular momentum from the gas to the bar itself. 
\cite{Berentzenetal2007} later suggested that the weakening of bars in discs with higher gas fractions is not due to the angular momentum transport between these two components, but rather due to the differences in bar buckling induced by the CMC.
CMCs that make up a few percent of the mass of the stellar disc, are thought to contribute to bar weakening and even destruction\footnote{It is worth noting that the masses and compactness of the CMCs required to completely destroy bars are not readily found in nature (see for example the discussion in \citealt{Athanassoulaetal2005}).}, and the more concentrated and massive the CMC, the larger this effect \citep{Hasanetal1993,Normanetal1996,Athanassoulaetal2005}. 
\cite{Athanassoulaetal2013} explored the formation of bars within galaxies with varying gas fractions and halos of different triaxiality. They found that galaxies that are more gas-rich stay axisymmetric for longer, and that once bars start to grow, they do so at a slower rate. 
Therefore, while gas appears to play a role in bar formation, it is unclear whether this occurs `directly' due to the exchange of angular momentum between the gas and the bar itself, or whether gas has an `indirect' effect on bar formation, for example through the formation of a CMC which forms when gas is pushed to the centre by the bar.

The aforementioned studies have largely been carried out using isolated N-body simulations. These lend themselves to exploring the internal evolution of galaxies, due to the higher resolution they offer, while they also enable one to isolate specific physical properties to study. However, it is becoming increasingly clear that the mechanisms responsible for bar formation and evolution cannot be decoupled from the cosmological context, since processes such as mergers (e.g. \citealt{Athanassoulaetal2017}), interactions (e.g. \citealt{Lokas2019}), external gas accretion (e.g. \citealt{BournaudCombes2002}), the physics of star formation and various feedback processes (e.g. \citealt{Irodotouetal2022}), as well as the assembly and properties of the dark matter halos themselves (e.g. \citealt{Collieretal2018}), all play a role in the formation and evolution of bars. Recent advances in terms of numerical resolution and modeling of the physics of baryons have made it possible to begin to use cosmological simulations to explore the internal dynamical evolution of galaxies both in `zoom-in simulations' (e.g. \citealt{ScannapiecoAthanassoula2012,Kraljicetal2012,Bucketal2018,Gargiuloetal2019,Fragkoudietal2020,Fragkoudietal2021,Merrowetal2024}) as well as in `large box' cosmological simulations (e.g. \citealt{Algorryetal2017,Peschkenetal2019,RosasGuevaraetal2020,Roshanetal2021,Frankeletal2022,Gargiuloetal2022,IzquierdoVillalbaetal2022,Lopezetal2024}). 
It is worth noting that all these simulations, which use different prescriptions for modelling the baryonic physics of galaxy formation and evolution, often lead to substantially different predictions for the fraction of bars (e.g. finding very few bars at low redshifts $z\sim0.25$; \citealt{Reddishetal2022}) as well as for their intrinsic properties (e.g. weak and transient bars; \citealt{Ansaretal2023}). As the galaxy formation (or `baryon' physics) models become more advanced, we can use the properties of bars to constrain these models, since bars act as a litmus test of whether the dynamics of observed discs are being adequately modelled in cosmological simulations. 

In this work we explore the properties of barred galaxies and how these evolve over cosmic history in the Auriga suite of magneto-hydrodynamical cosmological zoom-in simulations, to provide predictions for the properties of bars in Auriga at both low and high redshifts. We then explore the multi-dimensional parameter space of bar formation within the cosmological context, by examining the aforementioned parameters thought to be important for bar formation.
The paper is structured as follows: in Section \ref{sec:auriga}, we briefly describe the Auriga cosmological simulations, as well as the analysis tools used in this study. In Section \ref{sec:barprops} we present some of the global properties of barred galaxies in Auriga, and show how the length of bars evolves with redshift. In Section \ref{sec:drivers} we explore the different parameters that affect bar formation, and how these contribute to whether a galaxy will be barred or not in the cosmological context. In Section \ref{sec:discussion} we discuss some of the implications of our findings on the Tully Fisher relation, on bars at high redshift and on the multi-dimensional parameter space of bar formation. In Section \ref{sec:summary} we conclude and summarise our results.

\begin{figure*}
\centering
\includegraphics[width=0.98\textwidth]{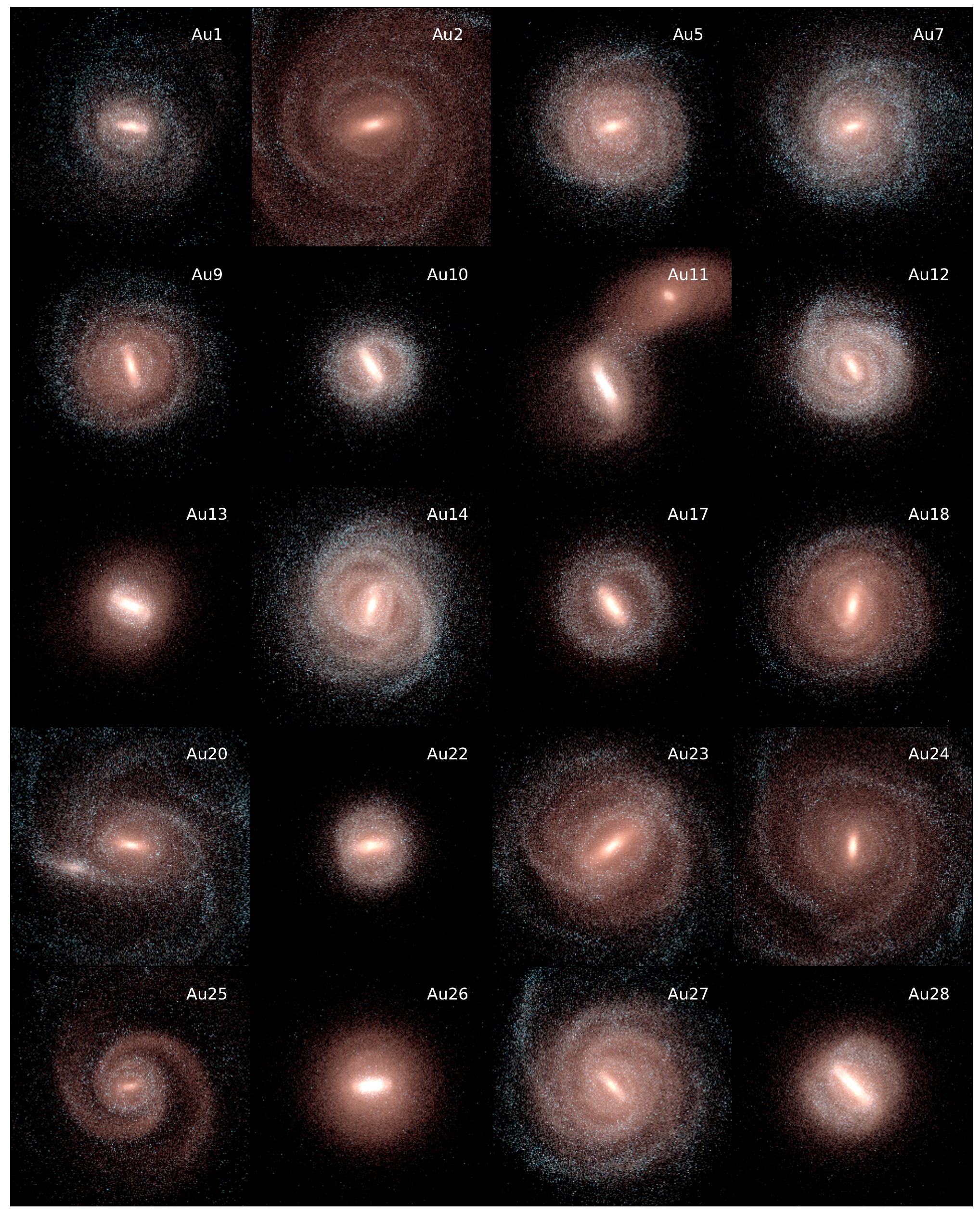}
\caption{Face-on RGB stellar light projections of the barred galaxies in the Auriga simulations at z=0. The panels have a size of 50\,kpc$\times$50\,kpc. } 
\label{fig:famport1}
\end{figure*}

\begin{figure*}
\ContinuedFloat
\centering
\includegraphics[width=0.98\textwidth]{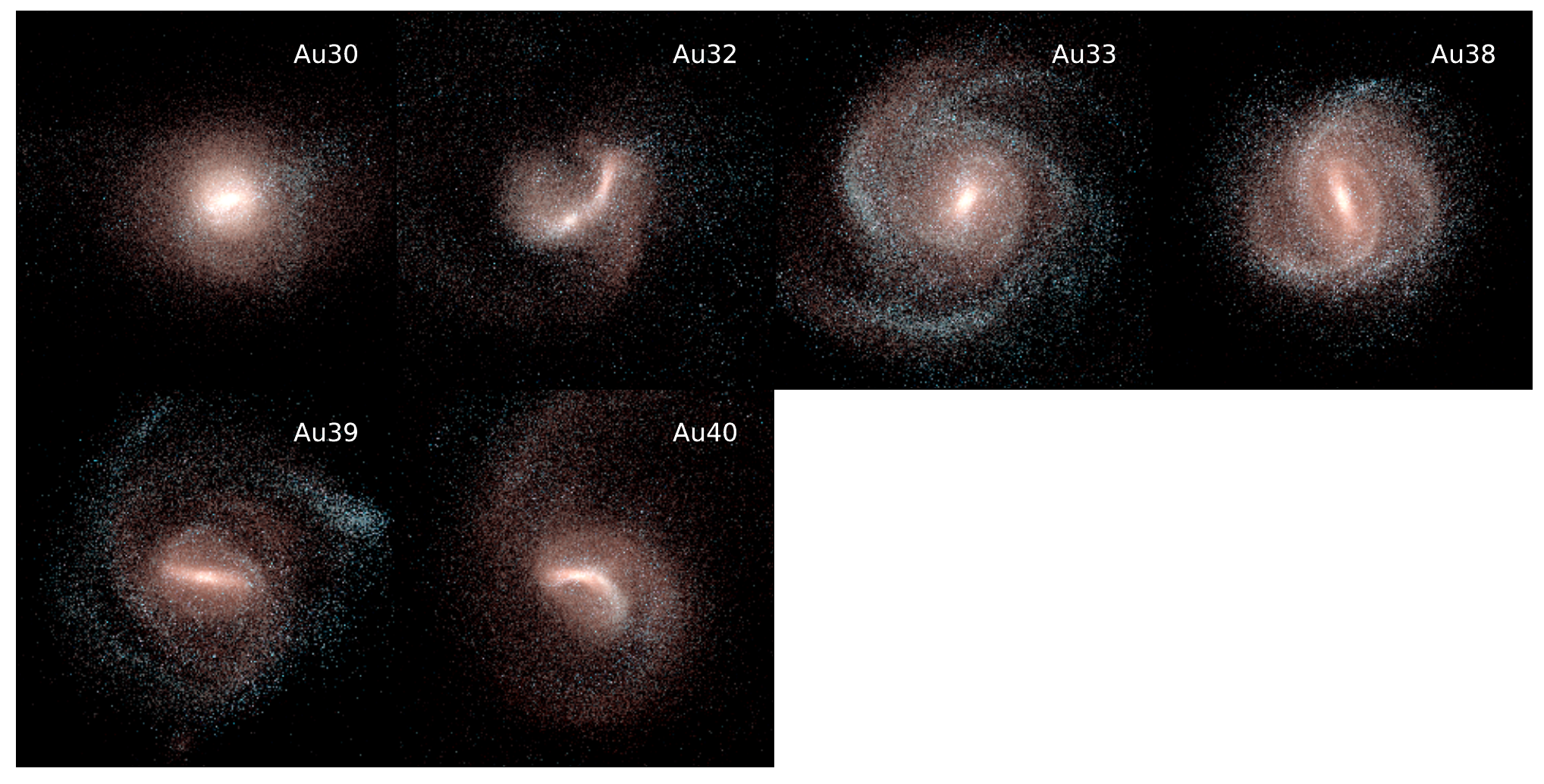}
\caption{Continued} 
\label{fig:famport2}
\end{figure*}


\section{The Auriga Simulations}
\label{sec:auriga}

\begin{figure}
\centering
\includegraphics[width=0.45\textwidth]{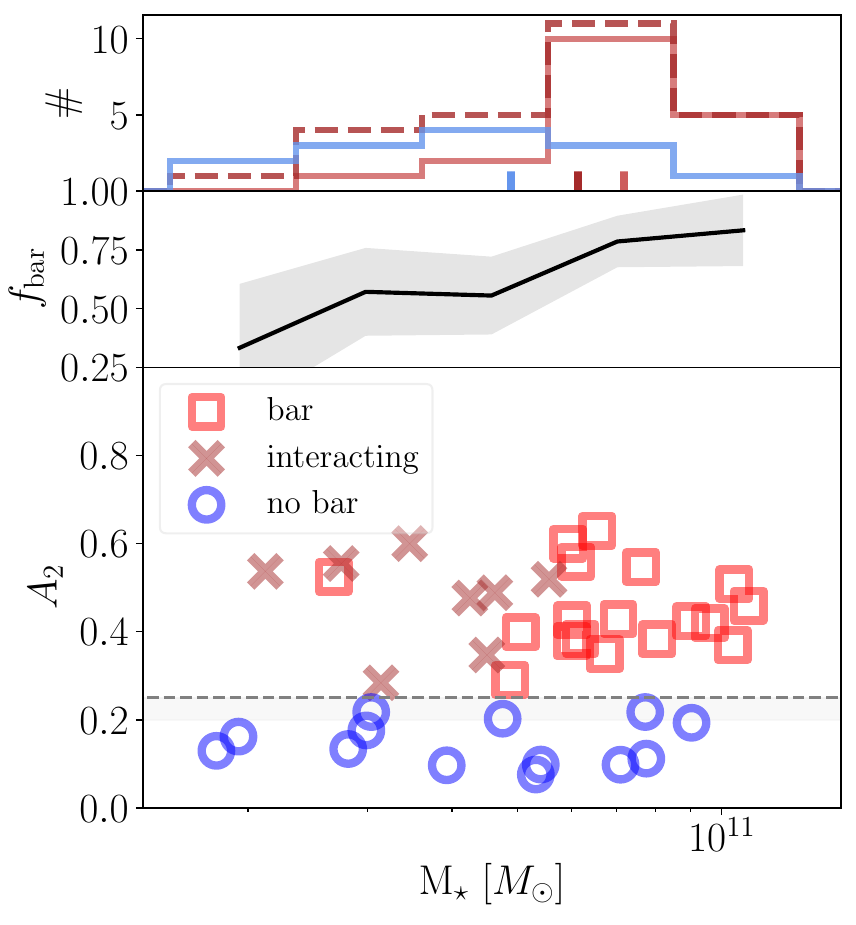}
\caption{\emph{Top:} Distribution of barred (red), unbarred (blue) and barred+interacting (dashed brown) galaxies in our sample in stellar mass. \emph{Middle:} Fraction of barred galaxies as a function of stellar mass. The shaded area is the uncertainty on the fractions assuming binomial statistics, given by $\sigma = \sqrt{f_{\rm bar}(1-f_{\rm bar})/N}$, where $N$ is the number of galaxies.
\emph{Bottom:} Bar strength versus stellar mass for the 40 galaxies in the Auriga suite of simulations. The horizontal dashed line indicates a bar strength of $A_2$=0.25, above which we consider the galaxies to be barred. Squares indicate barred galaxies, blue circles indicate unbarred galaxies, and crosses indicate barred galaxies which are undergoing an interaction at $z\sim0$ (see the text for more details).} 
\label{fig:summarysample}
\end{figure}

\begin{figure*}
\centering
\includegraphics[width=0.98\textwidth]{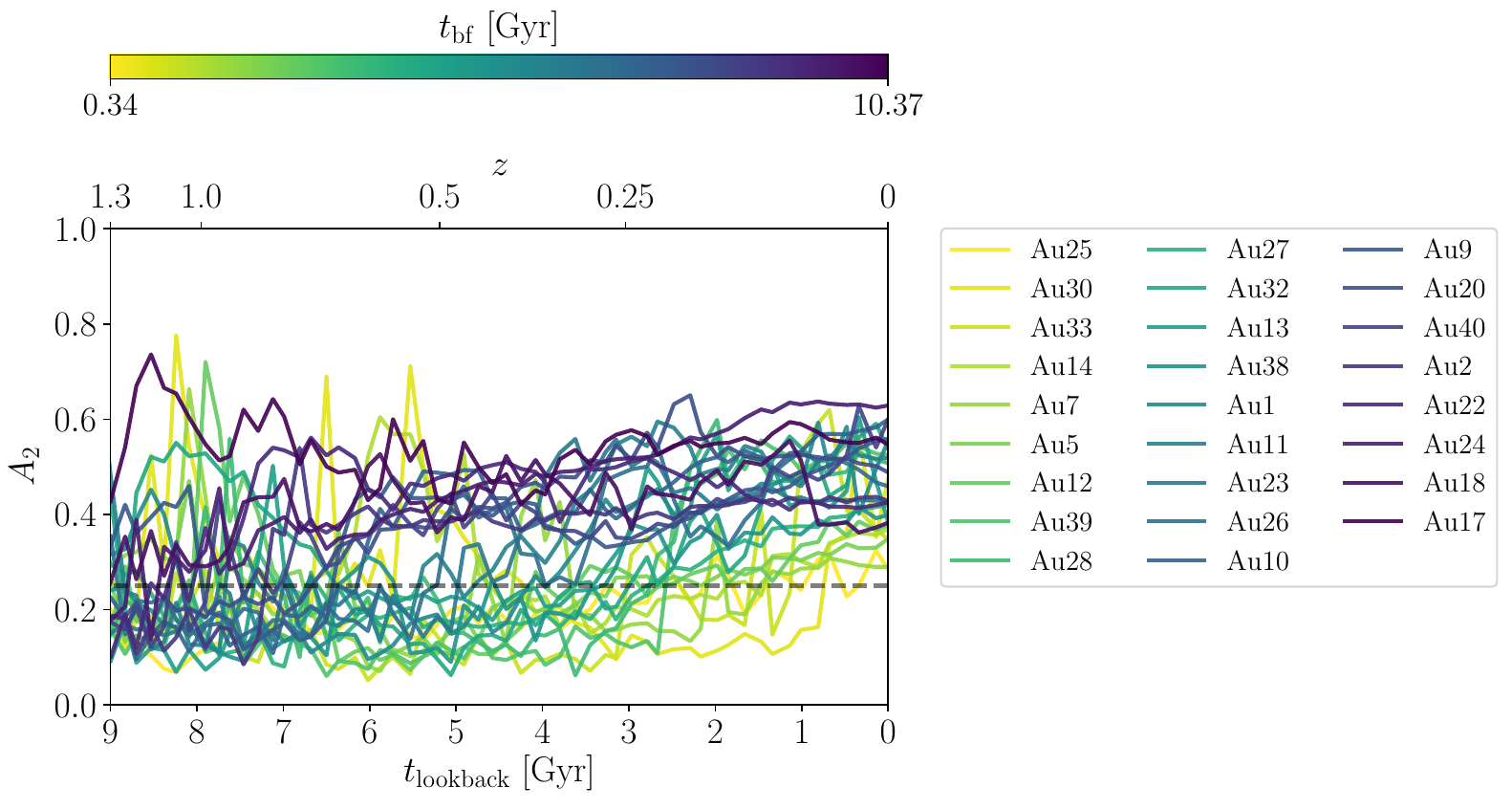}
\caption{The $m=2$ Fourier mode of the surface density as a function of lookback time for galaxies which are barred at $z=0$. The lines are sorted by the formation time of the bar (dark purple: bar forms at larger lookback time, light yellow: bar forms close to $z=0$). The galaxies that end up barred at $z=0$ develop their bars at various times in the simulation, some forming early on and staying barred, while others have bars that develop only at late times.} 
\label{fig:summarya2}
\end{figure*}

\begin{figure*}
\centering
\includegraphics[height=0.35\textwidth]{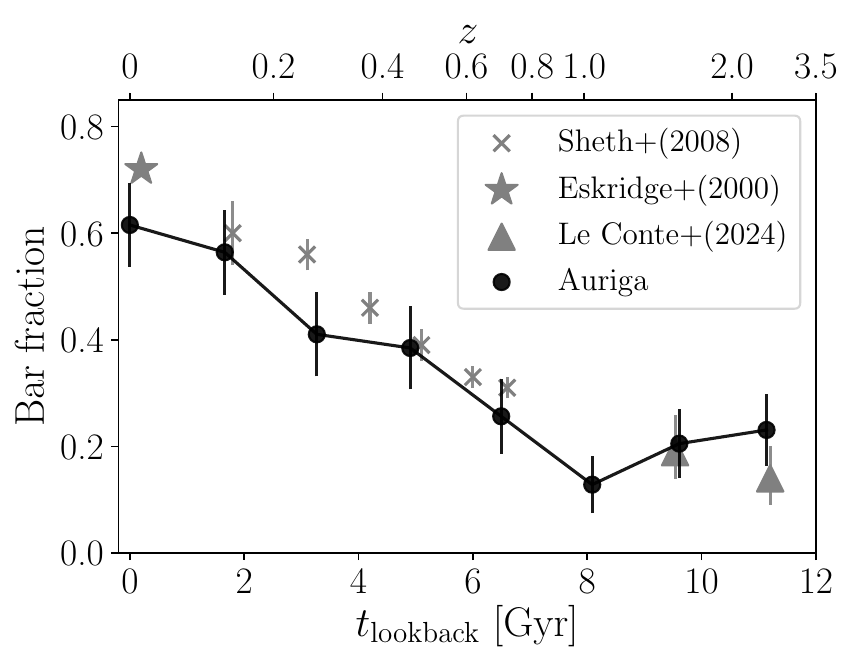}
\includegraphics[height=0.35\textwidth]{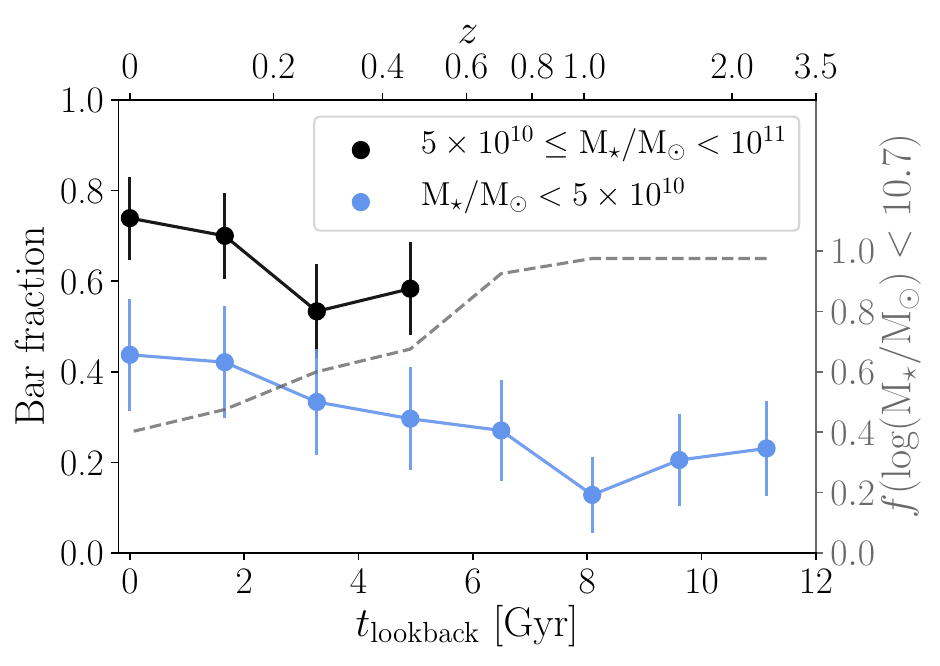}
\caption{Statistical properties of the entire sample of 40 Auriga galaxies. \emph{Left:} Bar fraction as a function of redshift (top axis) and lookback time (bottom axis) for the suite of Auriga simulations, compared to observations from \citet{Eskridgeetal2000}, \citet{Shethetal2008} and \citet{LeConteetal2024}. \emph{Right:} Bar fraction as a function of redshift for galaxies in two stellar masses bins, as indicated in the legend. The right axis and grey line indicate the fraction of galaxies in the low mass bin. We note that in these plots, time is plotted in the opposite direction from the other plots in the paper, in order to match the convention used in observational studies. We find a decreasing bar fraction with redshift in the Auriga simulations, and that more massive galaxies host more bars at all the times we are able to explore.} 
\label{fig:barfrac}
\end{figure*}

\begin{figure}
\centering
\includegraphics[width=0.4\textwidth]{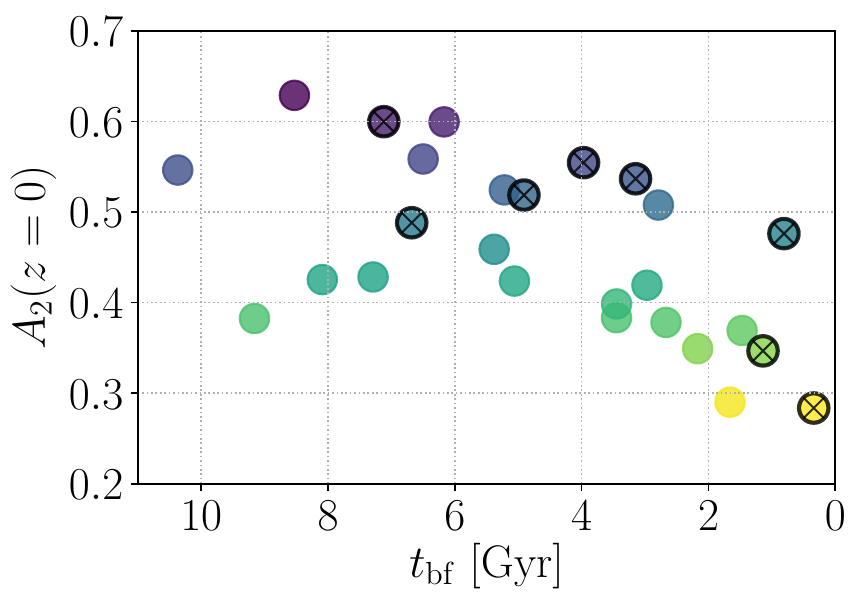}
\includegraphics[width=0.4\textwidth]{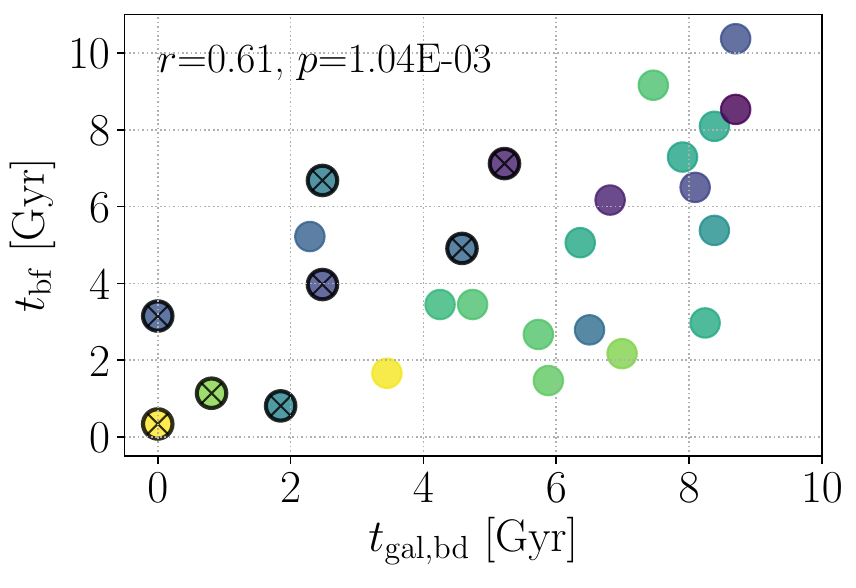}
\includegraphics[width=0.4\textwidth]{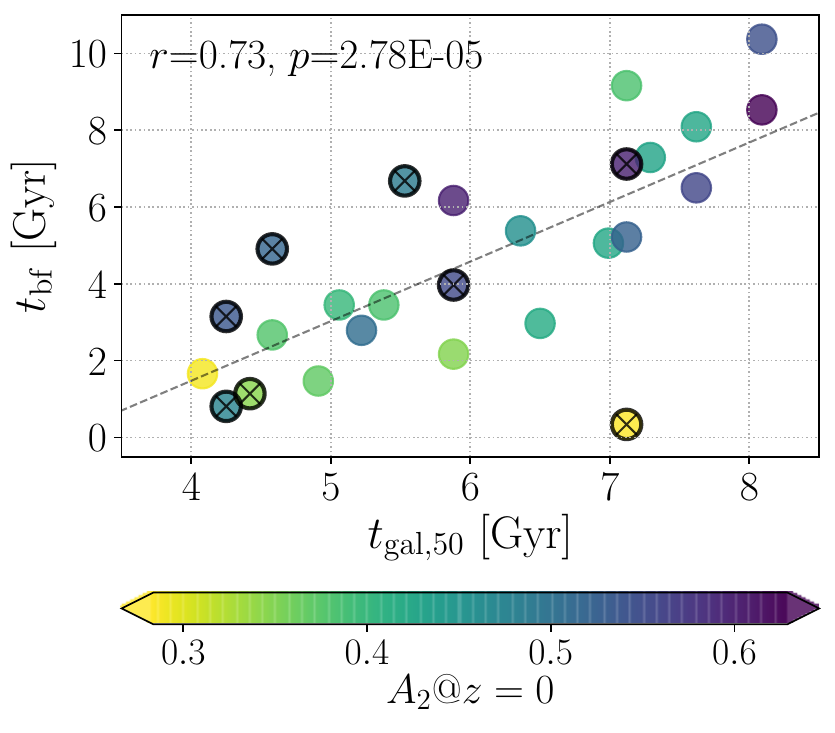}
\caption{\emph{Top:} Bar strength at $z=0$ versus the age of the bar. The colour-coding gives the bar strength (used in subsequent plots). The points outlined in black denote the `interacting' sample (see text and Table \ref{tab1}). 
\emph{Middle:} Bar formation time as a function of the time at which the galaxy becomes baryon-dominated. We indicate the Pearson coefficient, $r$, and the p-value for the entire sample of galaxies in the top left.
\emph{Bottom:} Bar formation time as a function of the time at which the galaxy assembles 50\% of its stellar mass. The dashed line indicates the linear regression fit for the non-interacting galaxies (see text).}
\label{fig:bars_vs_fbd}
\end{figure}

\begin{figure*}
\centering
\includegraphics[width=0.99\textwidth]{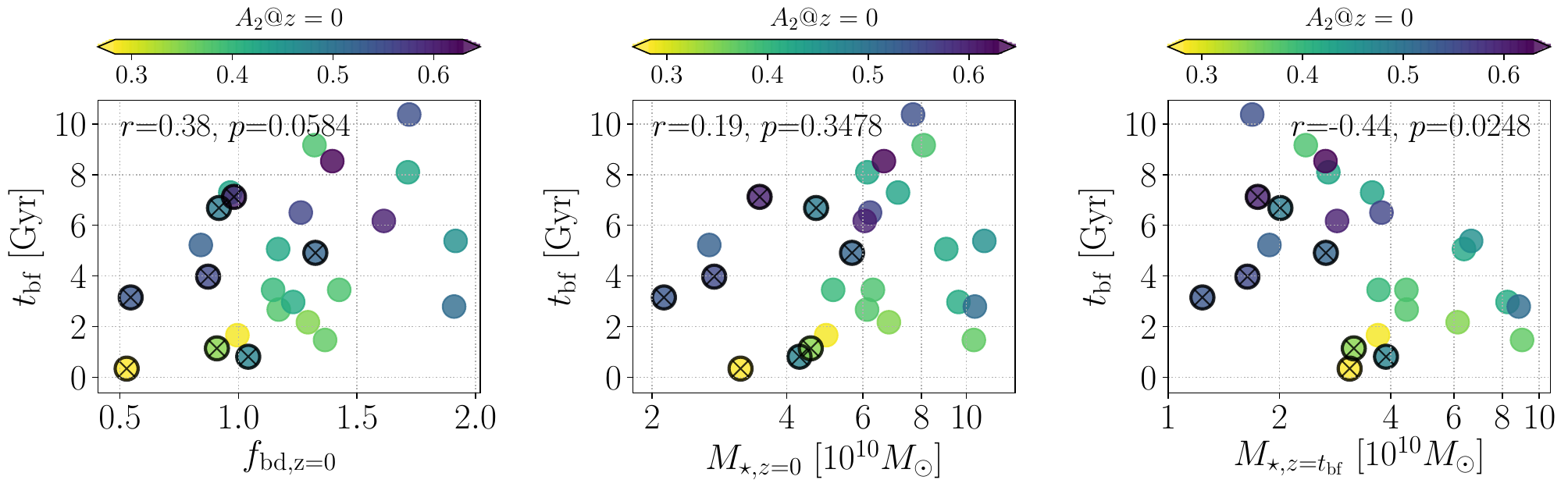}
\caption{Bar formation time as a function of various parameters. Black circles denote the interacting galaxies and the Pearson coefficient, $r$, and p-value are indicated at the top of each panel: \emph{Left:} Bar formation time as a function of the stellar-to-dark matter ratio inside 5\,kpc at $z=0$. \emph{Middle:}  Bar formation time as a function of the stellar mass of the galaxy at $z=0$. \emph{Right:} Bar formation time as a function of the stellar mass of the galaxy at the time of bar formation. There is a mild correlation between the bar formation time and the stellar-to-dark matter ratio at $z=0$, while there is no significant correlation between bar formation time and stellar mass at $z=0$ (i.e no obvious signs of mass downsizing within this mass range). There is a trend of decreasing stellar mass for earlier bar formation time (right panel), as expected, since galaxies at higher redshifts have lower stellar masses.} 
\label{fig:barformgalform}
\end{figure*}

\begin{figure}
\centering
\includegraphics[width=0.45\textwidth]{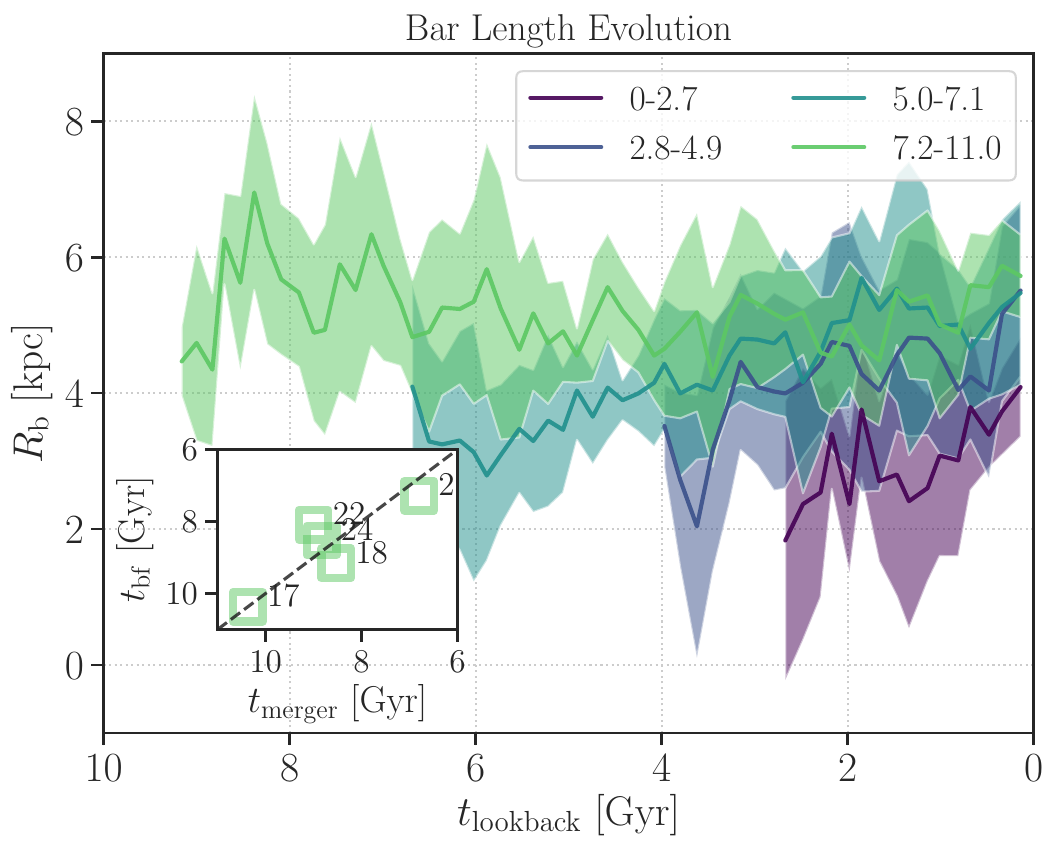}
\caption{Bar length as a function of time for galaxies that form bars between four different time-frames, as indicated by the figure inset. The solid lines show the average values of the bar length and the shaded region shows the $1\sigma$ dispersion. We see that bars that form at high redshifts tend to already form long, while bars forming at low redshifts form short and grow longer with time. The inset compares the time of significant merging events to the time of bar formation for the oldest bars, with the dashed line indicating the 1-to-1 line. The oldest bars in our sample seem to form around the time of a significant merger event.} 
\label{fig:barlength_vtime}
\end{figure}

\begin{figure*}
\centering
\includegraphics[width=0.99\textwidth]{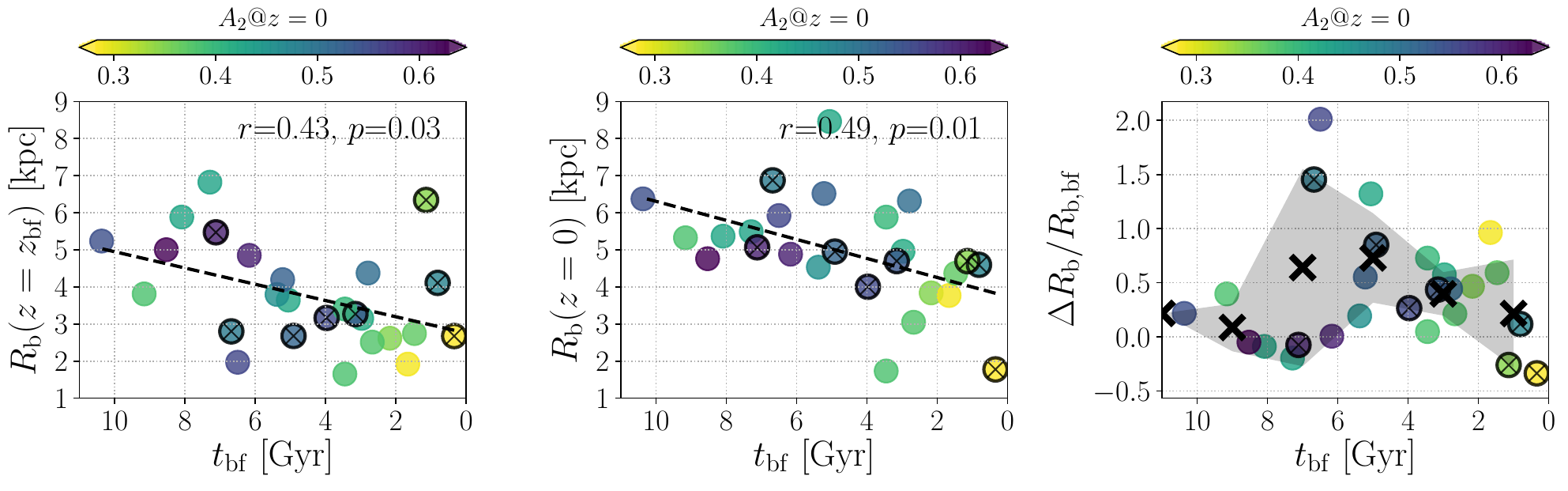}
\caption{Bar length as a function of the formation time of the bar for the entire barred sample (interacting galaxies have black circles). \emph{Left:} The bar length at the time of bar formation, as a function of bar formation time. \emph{Middle:} The bar length at $z=0$ for these same galaxies, as a function of the bar formation time. \emph{Right:} The relative change in the bar length as a function of bar formation time. The crosses indicate the average value for a given age bin, and the shaded region shows the 1$\sigma$ spread. Bars that form at high redshifts do not change their lengths much because they form long already, while the youngest bars have had limited amount of time to grow their bar length.} 
\label{fig:barlength}
\end{figure*}

\begin{figure*}
\centering
\includegraphics[width=0.99\textwidth]{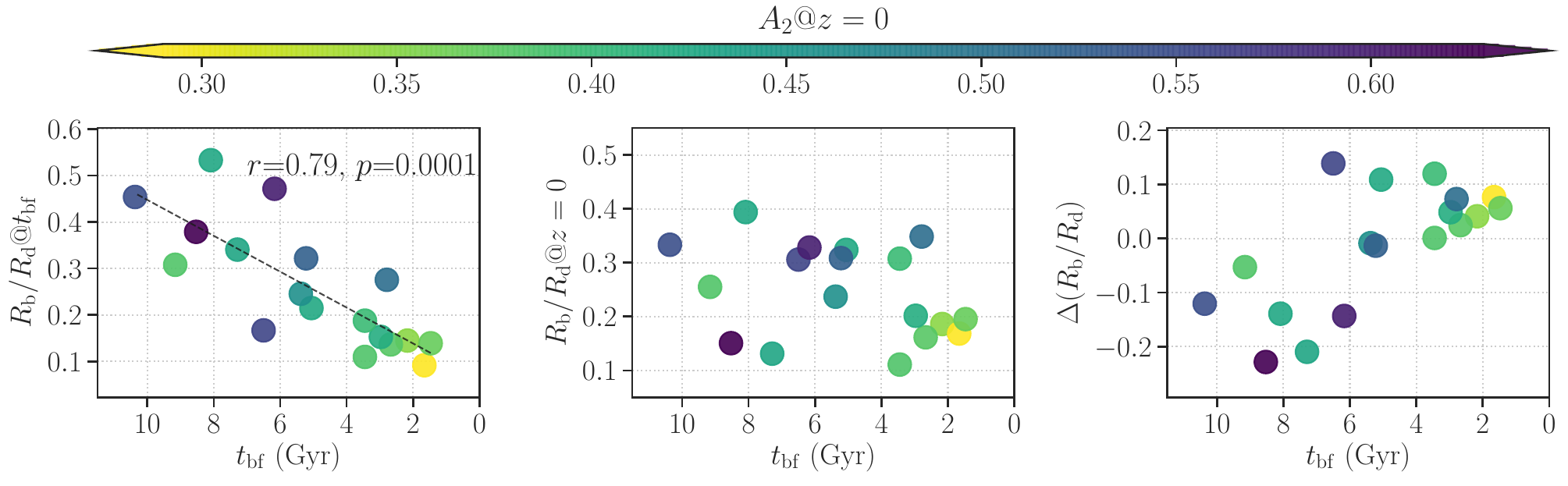}
\caption{The bar length-to-disc size ratio as a function of the formation time of the bar for the non-interacting galaxies. \emph{Left:} The ratio at the time of bar formation; \emph{Middle:} The ratio at $z=0$; \emph{Right:} The change in the ratio from bar formation time to $z=0$. We see that the ratio of bar length-to-disc size is larger at the time of bar formation for early forming bars, and decreases at $z=0$ due to the increasing disc size.} 
\label{fig:bardiscratio}
\end{figure*}

We explore the formation and evolution of bars in the Auriga simulations which are a suite of cosmological magneto-hydrodynamical zoom-in simulations of galaxies which span a range of halo masses at $z=0$ between $M_{\rm 200} = 0.5\times10^{12} -2\times 10^{12}\rm{M}_{\rm \odot}$ \citep{Grandetal2017,Grandetal2019}.
The simulations are run from redshift $z=127$ to $z=0$ with cosmological parameters $\Omega_{\rm m}=0.307$, $\Omega_{\rm b}=0.048$ and $\Omega_{\rm \Lambda}=0.693$, and a Hubble constant of $H_{\rm 0} =100h\, \rm km\,s^{-1}\,\rm Mpc^{-1}$,where $h=0.6777$ \citep{Planck2014XVI}. 
The simulations are performed with the magnetohydrodynamic code {\sc AREPO} \citep{Springel2010,Pakmoretal2016}, with a comprehensive galaxy formation model (see \citealt{Vogelsbergeretal2013, Marinaccietal2014a,Grandetal2017}, for more details) which includes primordial and metal line cooling, a prescription for a uniform background ultraviolet field for reionization (completed at $z=6$), a subgrid model for star formation, stellar evolution and feedback, magnetic fields, as well as black hole seeding, accretion and feedback. 
The dark matter particles have  a mass of $\sim 4\times10^5\rm{M}_{\odot}$ and the stars and gas have a mass resolution $\sim 5\times10^4\rm{M}_{\odot}$. The physical softening of collisionless particles grows with time and corresponds to a fixed comoving softening length of 500\,$h^{-1}$\,pc, while the maximum physical softening allowed is 369\,pc (see \citealt{Poweretal2003} for a criterion on how to choose softening parameters). The physical softening for the gas cells is scaled by the gas cell radius with a minimum limit of the softening equal to that of the collisionless particles. 

Star formation and stellar feedback are modelled as follows: if a given gas cell is eligible for star formation, it is converted (according to the \citealt{Chabrier03} initial mass function) either into a star particle -- in which case it represents a single stellar population of a given mass, age and metallicity -- or into a site for SNII feedback, i.e. a `wind' particle. 
Wind particles are launched in a random direction with a velocity that scales with the 1-D local dark matter velocity dispersion (see \citealt{Grandetal2017} for more details). The wind metal content is determined by the initial metallicity of the gas cell from which the wind particle originated, i.e. it is `loaded' with $\eta=0.6$ of the total metals of the parent gas cell. For the stellar particles, the mass loss and metal enrichment from SNIa and AGB stars is modelled and the mass and metals are distributed among nearby gas cells with a top-hat kernel.
We track a total of nine elements: H, He, C, O, N, Ne, Mg, Si, and Fe.

By $z=0$, the simulations form disc-dominated star-forming galaxies with flat rotation curves that reproduce a range of observed scaling relations such as the Tully-Fisher relation \citep{Grandetal2017} and the size-mass relation of HI gas discs \citep{Marinaccietal2017}. They also form instabilities in the discs such as bars and boxy/peanut bulges \citep{Fragkoudietal2020}, which have structural properties \citep{BlazquezCaleroetal2019} and pattern speeds \citep{Fragkoudietal2021} in agreement with those of observed bars, and mainly consist of so-called pseudo-bulges \citep{Gargiuloetal2019}, reproducing what is found for disc galaxies in the local Universe (e.g. \citealt{KormendyKennicutt2004,Gadotti2009}).  

We note that the full simulation output data for the Auriga simulations has been made publicly available -- see \citet{Grand2024} for more details. Various data products on barred galaxies in Auriga, collected in this paper, are now also available with the public data release\footnote{The data can be found at \url{https://wwwmpa.mpa-garching.mpg.de/auriga/}}.

\subsection{Barred galaxy sample}

In this work, we focus on the galaxies which at $z=0$ have halo masses in the range of $M_{\rm 200}$ = $0.5 \times 10^{12}-2 \times 10^{12}\rm{M}_{\odot}$.
This includes the 30 original Auriga halos presented in \cite{Grandetal2017} (Au1-Au30), as well as the 10 slightly lower mass galaxies, first presented in \cite{Grandetal2019} (Au31-Au40). We exclude Au34 from all the analysis that follows as it has a low resolution dark matter particle within its virial radius. 

In what follows we focus most of the analysis on the barred galaxies in the sample, which can be seen in Fig.~\ref{fig:famport1}. We determine whether a galaxy is barred by examining the bar strength, $A_2$, which we define as the $m=2$ Fourier mode of the stellar surface density at $z=0$ (see the next subsection for more details on how the bar strength is obtained). Our `barred' sample includes all the galaxies which at $z=0$ have $A_2\geq0.25$. This results in 26 out of the 39 halos explored here. The other 13 galaxies are classified as the unbarred sample. For parts of the analysis that follow, we further label the galaxies that undergo an interaction at $z\sim0$ with a companion. These are labeled as `interacting' galaxies and are identified by visually inspecting for disturbances in the stellar distribution of the galaxies within a lookback time of $t_{\rm lb}= 500$\,Myr. These samples are summarised in Table \ref{tab1} (see column `Classification') as well as in Fig.~\ref{fig:summarysample}, where we plot the bar strength $A_2$ as a function of stellar mass (bottom panel), the bar fraction versus stellar mass (middle panel) and the mass distribution of the three subsamples  (unbarred -- blue, barred without interacting -- red, and barred including interacting -- dashed brown).


\subsection{Analysis}
\label{sec:analysis}

Here we describe how we obtain the bar-related parameters for the galaxies in our sample, such as the bar strength, length and bar formation time.\\

\noindent \textbf{Bar strength, $A_2$:} To obtain the bar strength we first re-align the galaxy, such that the angular momentum vector of the stellar disc is parallel to the $z$-axis. We then select the stellar particles close to the disc, within a height of $|z|<0.8\,\rm{kpc}$ above and below the plane. The Fourier modes of the surface density are given by,

\begin{equation}
a_{\rm{m}}(R) = \sum^{N}_{i=1}  m_i \cos(\rm{m}\theta_i), \,\,\,\,\, \,\,\,\,\, m=0,1,2,...,
\end{equation}

\begin{equation}
b_{\rm{m}}(R) = \sum^{N}_{i=1} m_i \sin({\rm{m}}\theta_i),  \,\,\,\,\,\,\,\,\,\, {\rm{m}}=0,1,2,...
\end{equation}

\noindent where m denotes the Fourier mode, $m{_i}$ is the mass of particle $i$, $R$ is the cylindrical radius, $N$ is the total number of particles at that radius and $\theta_{i}$ is the azimuthal angle.
In order to obtain a single value for the bar strength we take the maximum value of the relative $m=2$ component within the inner 10\,kpc as,
\begin{equation}
A_2 = \rm max \left(\frac{\sqrt{ \left( a^2_2(R) + b^2_2(R) \right)} }{a_0(R)}\right).
\end{equation} 

For estimating the bar fraction in Fig.~\ref{fig:barfrac} we also visually inspect the galaxies, to ensure that when $A_2\geq0.25$, this is not due, for example, to a transient off-centering of the disc due to a merger (which can occur frequently at high redshifts). The aforementioned parameters and cuts used for the bar strength are applied such that our definition of `barred galaxy' agrees well with tests that were done in which the authors visually identified bars in the simulations. We note that slight changes in these parameters do not change the overall conclusions of this work.\\

\noindent \textbf{Bar formation time, $t_{\rm bf}$:} In what follows we define the bar formation time $t_{\rm bf}$ as the first lookback time at which $A_2$ becomes, and subsequently remains, larger than 0.25. In Fig.~\ref{fig:summarya2} we show the bar strength $A_2$ as a function of lookback time, for the barred galaxies in our sample, which we colour-code according to their bar formation formation time.\\

\noindent \textbf{Bar length, $R_{\rm bar}$:} There are various methods for estimating the length of the bar, from ellipse-fitting methods used widely in observational studies, to methods using the $m=2$ Fourier mode of the surface density, which are more commonly employed in numerical studies (see e.g. \citealt{Erwin2005,Gadotti2011,Hilmietal2020,Fragkoudietal2021,Petersenetal2023}). Here we use a method which is well-suited for estimating the bar length in an automated way for simulated galaxies, across a large number of snapshots, i.e. we use the radius at which $A_2$ falls below 70\% of its maximum value. The bar length at the time of bar formation can vary rapidly, in particular when the bar forms due to an interaction or after a merger. Therefore, when estimating the bar length at the time of formation, we take the average over three snapshots after $t_{\rm bf}$. We note that we have tested various methods for obtaining the bar length, such as a definition based on the constancy of the phase of the $m=2$ Fourier mode. We found this latter method to be more prone to noisy bar length estimates, hence why we use the former method here (see also Figure B2 of \citealt{Fragkoudietal2021}).  \\

\noindent \textbf{Disc size, $R_{\rm d}$:} We define the size of the disc for the galaxies in our sample as the radius at which the stellar surface density drops below 1$\rm{M}_{\odot}/\rm pc^2$. \\

\noindent \textbf{Time at which the galaxy becomes baryon-dominated, $t_{\rm gal,bd}$:} We define the time at which the galaxy becomes baryon-dominated, $t_{\rm gal,bd}$, as the time at which the stellar mass begins to contribute at least 80\% of the total mass within 5\,kpc.
In what follows, when referring to the stellar mass of a galaxy, we use the mass within 10\% of the virial radius of the galaxy, unless otherwise noted.\\

\noindent \textbf{The galaxy assembly time, $t_{\rm gal,50}$:} is defined as the time at which the galaxy assembles 50\% of its stellar mass within a radius of 20\,kpc. \\

\noindent \textbf{$M_{\rm \star}$:} In what follows, we calculate the stellar mass of the simulated galaxy as the mass within 10\% of the virial radius, unless otherwise stated.

\section{Bar properties \& evolution}
\label{sec:barprops}

We start by exploring some statistical and global properties of bars in our sample (Section \ref{sec:globalprops}), such as the fraction of bars across stellar mass and redshift, the formation time of bars and their longevity. We then look into one of the most important bar properties, the bar length, its relation to the size of the disc, and how these evolve in time (Section \ref{sec:barlength}).

\subsection{Global Properties of Barred Galaxies}
\label{sec:globalprops}

\begin{table*}
\begin{center}
\begin{tabular}{ p{0.7cm}|p{1.3cm}|p{0.7cm}|p{1.4cm}|p{1.6cm} |p{1.4cm} |p{1.6cm} |p{1.4cm} |p{1.6cm} |p{1.5cm}    }
\hline
 Halo & $A_2(z=0)$ & $t_{\rm bf}$ & $R_{\rm b} (z=0)$ & $R_{\rm b} (z=z_{\rm bf})$ & $R_{\rm d} (z=0)$ & $R_{\rm d} (z=z_{\rm bf})$ & $M_{\rm \star} (z=0)$ & $M_{\rm \star} (z=z_{\rm bf})$ & Classification \\
   &  & [Gyr] & [kpc] & [kpc] & [kpc]  & [kpc]  & [$10^{10}\rm{M}_{\odot}$] & [$10^{10}\rm{M}_{\odot}$] & \\
 \hline
Au1  &  0.55  &  3.97  &  4.00  &  3.16  &  18.38  &  8.38 &   2.74   &   1.63    &  bar, int  \\
Au2  &  0.43  &  7.29  &  5.49  &  6.82  &  41.88  &  20.04 &   7.04   &   3.55    &  bar   \\
Au3  &  0.11  &     &   &     &  32.12  &    &   7.75   &         &  unbarred   \\
Au4  &  0.10  &     &   &     &  20.38  &    &   7.09   &         &  unbarred   \\
Au5  &  0.35  &  2.17  &  3.84  &  2.62  &  20.62  &  18.04 &   6.72   &   6.02    &  bar   \\
Au6  &  0.20  &     &   &     &  27.62  &    &   4.75   &         &  unbarred   \\
Au7  &  0.29  &  1.66  &  3.77  &  1.92  &  22.38  &  20.88 &   4.87   &   3.68    &  bar   \\
Au8  &  0.18  &     &   &     &  27.88  &    &   2.99   &         &  unbarred   \\
Au9  &  0.56  &  6.50  &  5.92  &  1.97  &  19.38  &  11.79 &   6.10   &   3.76    &  bar  \\
Au10  &  0.60  &  6.17  &  4.88  &  4.85  &  14.88  &  10.29 &   5.94   &   2.85    &  bar   \\
Au11  &  0.52  &  4.91  &  4.95  &  2.68  &  20.38  &  15.21 &   5.56   &   2.66    &  bar, int  \\
Au12  &  0.38  &  2.67  &  3.05  &  2.51  &  18.88  &  18.38 &   6.01   &   4.39    &  bar    \\
Au13  &  0.38  &  3.45  &  1.73  &  1.65  &  15.62  &  15.04 &   6.19   &   4.38    &  bar    \\
Au14  &  0.37  &  1.47  &  4.37  &  2.74  &  22.38  &  19.71 &   10.4 &   8.98    &  bar   \\
Au15  &  0.10  &     &    &     &  23.12  &    &   3.93   &         &  unbarred   \\
Au16  &  0.10  &     &   &     &  36.88  &    &   5.41   &         &  unbarred   \\
Au17  &  0.55  &  10.37  &  6.37  &  5.24  &  19.12  &  11.54 &   7.61   &   1.68    &  bar   \\
Au18  &  0.38  &  9.16  &  5.33  &  3.81  &  20.88  &  12.38 &   8.04   &   2.35    &  bar   \\
Au19  &  0.08  &     &    &     &  25.88  &    &   5.32   &         &  unbarred  \\
Au20  &  0.49  &  6.68  &  6.88  &  2.80  &  28.38  &  15.88 &   4.63   &   2.01    &  bar, int  \\
Au21  &  0.22  &     &    &     &  24.38  &    &   7.72   &         &  unbarred   \\
Au22  &  0.43  &  8.09  &  5.37  &  5.88  &  13.62  &  11.04 &   6.02   &   2.70    &  bar   \\
Au23  &  0.42  &  5.06  &  8.46  &  3.64  &  26.12  &  16.96 &   9.02   &   6.27    &  bar   \\
Au24  &  0.63  &  8.53  &  4.76  &  5.00  &  31.62  &  13.21 &   6.55   &   2.65    &  bar   \\
Au25  &  0.28  &  0.34  &  1.77  &  2.67  &  26.12  &  25.12 &   3.14   &   3.08    &  bar, int  \\
Au26  &  0.46  &  5.38  &  4.53  &  3.80  &  19.12  &  15.46 &   11.0 &   6.55    &  bar   \\
Au27  &  0.42  &  2.97  &  4.96  &  3.15  &  24.62  &  20.62 &   9.61   &   8.21    &  bar  \\
Au28  &  0.51  &  2.79  &  6.32  &  4.37  &  18.12  &  15.88 &   10.4 &   8.80    &  bar   \\
Au29  &  0.19  &     &   &     &  18.62  &    &   9.03   &         &  unbarred    \\
Au30  &  0.48  &  0.81  &  4.60  &  4.11  &  18.62  &  19.38 &   4.25   &   3.86    &  bar, int  \\
Au31  &  0.13  &     &   &     &  19.88  &    &   1.80   &         &  unbarred, int  \\
Au32  &  0.54  &  3.15  &  4.70  &  3.27  &  18.88  &  21.12 &   2.12   &   1.24    &  bar, int  \\
Au33  &  0.35  &  1.14  &  4.70  &  6.34  &  23.12  &  19.21 &   4.50   &   3.16    &  bar, int  \\
Au35  &  0.16  &     &    &     &  19.38  &    &   1.94   &         &  unbarred, int  \\
Au36  &  0.22  &     &    &     &  20.88  &    &   3.04   &         &  unbarred, int  \\
Au37  &  0.13  &     &    &     &  25.38  &    &   2.81   &         &  unbarred   \\
Au38  &  0.40  &  3.45  &  5.88  &  3.40  &  19.12  &  18.12 &   5.05   &   3.69    &  bar   \\
Au39  &  0.52  &  5.22  &  6.52  &  4.20  &  21.12  &  13.04 &   2.68   &   1.87    &  bar   \\
Au40  &  0.60  &  7.12  &  5.07  &  5.47  &  21.88  &  11.46 &   3.46   &   1.74    &  bar, int  \\
 \hline
\end{tabular}
\caption{\label{tab1} Properties of barred galaxies: (1) Halo name; (2) $A_2$ the bar strength at $z=0$ for the 39 galaxies in this sample; (3) the lookback time of bar formation ($t_{\rm bf}$) in Gyr; (4) The length of the bar at $z=0$, (5) the length of the bar at the time of bar formation, (6) the size of the disc at $z=0$, (7) the size of the disc at the time of bar formation, (8) the stellar mass of the galaxy at $z=0$, (9) the stellar mass of the galaxy at the time of bar formation, (10) classification of the galaxy as barred, unbarred and whether it falls within our definition of interacting at $z=0$ (int; see text for more details).}
\end{center}
\end{table*}

\noindent \textbf{Bar Formation Time:}
In Fig.~\ref{fig:summarya2} we show the $m=2$ Fourier mode for the barred galaxies in our sample, as a function of time, colour-coded according to their bar formation time. We see that bars have a wide range of formation times (see also Table \ref{tab1}), with the oldest bar in our sample forming at a lookback time of $t_{\rm lb}=10.4$\,Gyr and the youngest bar forming at $t_{\rm lb}=0.3 \,$Gyr. It is worth emphasizing that, from $z\sim1-2$, bars in Auriga are long-lived structures, i.e. they do not dissolve and are not recurring. This is evident in Fig.~\ref{fig:summarya2} where we see that once a bar forms, i.e. once the bar strength is above $A_2>0.25$, it remains above 0.25 and does not go through cyclical periods of destruction and reformation. The single exception to this rule is Au36, whose bar is severely weakened by an interaction with another galaxy, resulting in it being defined as unbarred by $z=0$ (see Fig.~\ref{fig:a2nobar}; we will explore this system in more detail elsewhere).
\\

\noindent \textbf{Bar fraction:}
In Fig.~\ref{fig:barfrac} we explore the fraction of bars as a function of redshift (and lookback time) in our sample, comparing this to observations from \cite{Eskridgeetal2000}, \cite{Shethetal2008} and \cite{LeConteetal2024}. We find that $60\%$ of the galaxies in our sample at $z=0$ are barred\footnote{This is slightly lower than the fraction found in \cite{Fragkoudietal2020}, where we only included the 30 high mass Auriga halos, while here we use $A_2>0.25$ to identify bars, as opposed to 0.2 which was employed in our previous study.}, with the fraction of bars decreasing for higher redshifts and reaching about $\sim 20\%$ at $z\sim3$, in agreement with the aforementioned observations. Beyond $z\sim1$ there appears to be a slight increase in the bar fraction in our simulations (although not that this is within the uncertainties), indicating that some structures identified as bars at high redshifts might be transient.

The right panel of Fig.~\ref{fig:barfrac} shows the bar fraction as a function of lookback time in two stellar mass bins, with masses greater than, and smaller than $M_{\star}=5\times 10^{10} \rm{M}_{\odot}$. The right $y$-axis and grey line indicate the fraction of galaxies in the low mass bin. We find that the bar fraction is higher in the high mass bin at all lookback times at which these galaxies exist in our sample. This emphasizes that at all times, bars are more common in massive galaxies. These findings seem to be compatible with the mass downsizing hypothesis, i.e. that more massive galaxies form bars first (see for example \citealt{Shethetal2008}), but as we show below, this does not necessarily imply a straightforward correlation between the stellar mass of the galaxy and bar formation time. \\

\noindent \textbf{Mass Build-Up and Bar Formation Time}
In the top panel of Fig.~\ref{fig:bars_vs_fbd} we show the bar strength at $z=0$ as a function of the bar formation time. We find a weak correlation between the bar strength at $z=0$ and the age (or formation time) of the bar, indicating that older bars tend to be slightly stronger. In the second panel of the figure we show the bar formation time as a function of the time at which the galaxy becomes baryon dominated (see Section \ref{sec:analysis} for details). We see that there is a correlation between these quantities, suggesting that galaxies form a bar when the galaxy passes a threshold of baryon-dominance (and see also Section \ref{sec:baryondom}). 

In the bottom panel of Fig.~\ref{fig:bars_vs_fbd} we show the bar formation time as a function of the formation time of the galaxy (which we define as the time at which the galaxy assembles 50\% of its stellar mass). We find a strong correlation between the two parameters, indicating that younger bars tend to form in galaxies that assembled their mass later. We note that the yellow point which is an outlier in this relation (Au25) has an underestimated bar age due to the method we use for obtaining the bar formation time. As discussed in Sec.~\ref{sec:analysis}, the bar age is taken as the last time $A_2$ becomes larger than 0.25. This galaxy has a bar strength that oscillates around the cut-off value of 0.25, starting at around $t_{\rm lb}\sim3.25$\,Gyr. The last time it crosses above 0.25 is at $t_{\rm lb}=0.34$\,Gyr -- making this the bar age according to our definition. However, there is a clearly formed -- albeit weak -- bar at earlier times already at $t_{\rm lb}\sim3.25$\,Gyr. The dashed grey line indicates a linear fit to the sample with a slope of 1.6 and intercept of -4.7\,Gyr.

In Figure \ref{fig:barformgalform} we explore whether the time of bar formation is correlated to the stellar mass of the galaxy at $z=0$, as one might naively expect in a mass downsizing scenario. In the middle panel we show the formation time of the bar as a function of the stellar mass of the galaxy at $z=0$, in which we do not find a correlation. Nor do we find a correlation between bar formation time and the stellar mass of the galaxy at the time of bar formation, as is shown in the right panel of Figure \ref{fig:barformgalform} -- indeed we find a weak anti-correlation between these quantities. This can be understood since bars that form at higher redshifts will form in galaxies that have not yet fully assembled their stellar mass. 
On the other hand, we see that the formation time of the bar correlates (albeit weakly) with how much the stellar component dominates over the dark matter component within 5\,kpc at $z=0$, as can be seen in the left panel of Figure \ref{fig:barformgalform}. This indicates that at $z=0$, we can relate how old a galaxy's bar is, with how baryon-dominated that galaxy is. It is worth noting the scatter in this relation, which increases at older ages, likely due to effects related to mergers which, as we will see below, can also act as triggers of bar formation.

\subsection{Evolution of Bar Properties}
\label{sec:barlength}

In this section we explore the evolution of bar properties in Auriga as a function of time, focusing in particular on the length of bars and the ratio of bar length to disc size.

In Figure \ref{fig:barlength_vtime} we show the evolution of bar length as a function of time. In particular, we group together bars according to their formation time, i.e. we select bars that form between lookback times of $t_{\rm lb}=0-2.7$\,Gyr ago (dark purple), $t_{\rm lb}=2.8-4.9$\,Gyr (dark blue), $t_{\rm lb}=5-7.1$\,Gyr (dark green) and $t_{\rm lb}=7.2-11$\,Gyr (light green), as indicated by the legend. These bins are chosen such that they split the sample into 4 groups with approximately the same number of galaxies. We find that bars that form at low redshifts, tend to form shorter bars which subsequently grow longer over time. 

On the other hand, bars that form at higher redshifts (e.g. around $z\sim 1$ -- see the green line) tend to already be quite long when they form, and their length does not evolve much over the galaxy's lifetime. We hypothesize that this is due to the fact that galaxies that form bars at high-$z$ are triggered by a merger, and therefore form long and ``saturated'', i.e. they are not able to grow by exchanging more angular momentum. Indications of this can be seen in the inset of Fig.~\ref{fig:barlength_vtime}, where we show the time of one of the three most massive mergers in the galaxy, versus the time of bar formation. We see that the galaxies that form bars at high redshift all do so around the time of one of their most massive mergers. 
The growth of bars is mediated through the exchange of angular momentum between the inner disc and the outer disc and halo, with the disc acting as an `emitter' and the halo acting as a `sink' of angular momentum (see e.g. \citealt{Athanassoula2003}). 
We speculate that if a bar forms through a merger/interaction, the torques induced in this process might reduce the amount of angular momentum able to be exchanged. This could be because of the torques induced in the disc which can trigger the bar to form in the first place (see e.g. \citealt{Lokas2018,Merrowetal2024}) or it could be due to the injection of angular momentum in the halo via the interaction/merger, which therefore means the halo is able to absorb less angular momentum. These processes could therefore lead to the angular momentum exchange being `saturated', which can impede the further growth of the bars. While there are hints of this in the simulations (e.g. the fact that the bars forming at high redshifts all form around the time of a significant merger, and that they do not grow longer in time), this hypothesis would need to be investigated in more detail, which we defer to future work.

We quantify these trends further in Fig.~\ref{fig:barlength}, where we plot the bar length of galaxies as a function of the formation time of the bar, in the left panel at the time of bar formation, and in the middle panel at $z=0$. In the right panel we show the relative change in the bar length from the time of bar formation until $z=0$ for each galaxy. This quantifies the bar growth for the various galaxies, reinforcing the fact that the oldest bars undergo less growth -- as they already form long -- while the bars that undergo the most growth are those of intermediate ages. The youngest bars also grow in time, but as they have less time to evolve than the intermediate age bars, their relative growth is limited.

Given that galaxies tend to have smaller discs at higher redshifts (e.g. \citealt{vanderWeletal2014,Ormerodetal2024}), it might be somewhat surprising that bars that form early on in the Universe in our simulations have longer bars. This would imply that the ratio of bar length-to-disc size changes as a function of redshift. We explore the ratio of bar length to disc size in Fig.~\ref{fig:bardiscratio}; we focus here on the non-interacting barred galaxies, since interacting galaxies have disturbed discs and as such the concept of the disc size becomes ambiguous. In the left panel of the figure we show the ratio of bar length-to-disc size \emph{at the time of bar formation} for each of the galaxies, as a function of bar formation time. We find a trend in which this ratio is higher for galaxies whose bars form at higher redshifts. We therefore see that when bars form at high redshifts they extend over a larger range of the disc. In the middle panel we show this ratio at $z=0$, and in the right panel we show the change in the ratio from the time of bar formation until $z=0$. In the middle panel it is clear that the bar length-to-disc size relation flattens out by $z=0$. By examining the middle and right panels of the figure, we see that the ratio decreases for the older bars, due to the fact that the disc size increases for these galaxies, while the bar length doesn't increase significantly (as shown in Figs. \ref{fig:barlength_vtime} and \ref{fig:barlength}). On the other hand, for the younger bars the ratio increases at $z=0$, which is due to the increase in the bar length for intermediate and young bars over time. Therefore, when observing bars at higher redshift, we might expect to find a higher bar-to-disc size ratio than in the local Universe.

\section{Drivers of bar formation}
\label{sec:drivers}

In this section we explore the various factors that are thought to be important for the formation of bars -- namely the baryon-dominance (\ref{sec:baryondom}), accreted bulge component (\ref{sec:exsitu}), gas fraction (\ref{sec:gasfrac}) and the Toomre Q parameter (\ref{sec:veldisp}) -- examining the role these play in bar formation within the cosmological context.

\subsection{Baryon dominance}
\label{sec:baryondom}

\begin{figure*}
\centering
\includegraphics[width=0.99\textwidth]{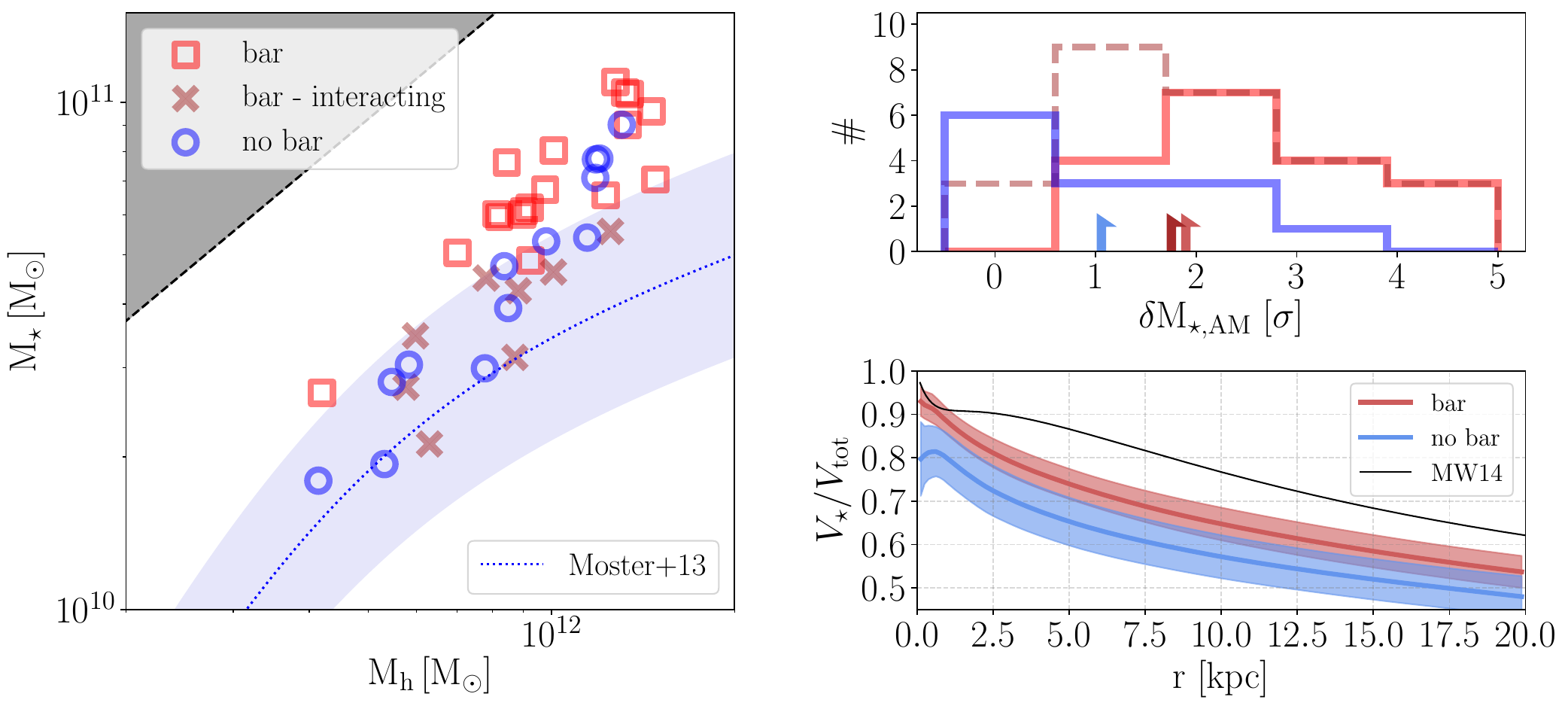}
\caption{\emph{Left:} The loci of galaxies in our sample in the stellar mass - halo mass plane at $z=0$. We indicate whether the galaxy is unbarred (blue circle), barred (red square) or barred and undergoing an interaction at $z=0$ (cross). The abundance matching relation from \citet{Mosteretal2013} is indicated with the blue line and shaded region. \emph{Top Right:} The distance of galaxies from the abundance matching relation for barred (red), unbarred (blue) and barred+interacting (dashed brown) galaxies. \emph{Bottom Right:} The baryon-dominance as a function of radius of the barred (without interacting galaxies; red) and unbarred (blue) galaxies in our sample, as indicated by $V_{\star}$/$V_{\rm{tot}}$. We compare these to the Milky Way model from \citet{BovyRix2013} (black). While barred galaxies are on average more baryon-dominated than unbarred galaxies in Auriga, both are less baryon-dominated than the MW14 model for the Milky Way. } 
\label{fig:amall}
\end{figure*}

\begin{figure}
\centering
\includegraphics[width=0.48\textwidth]{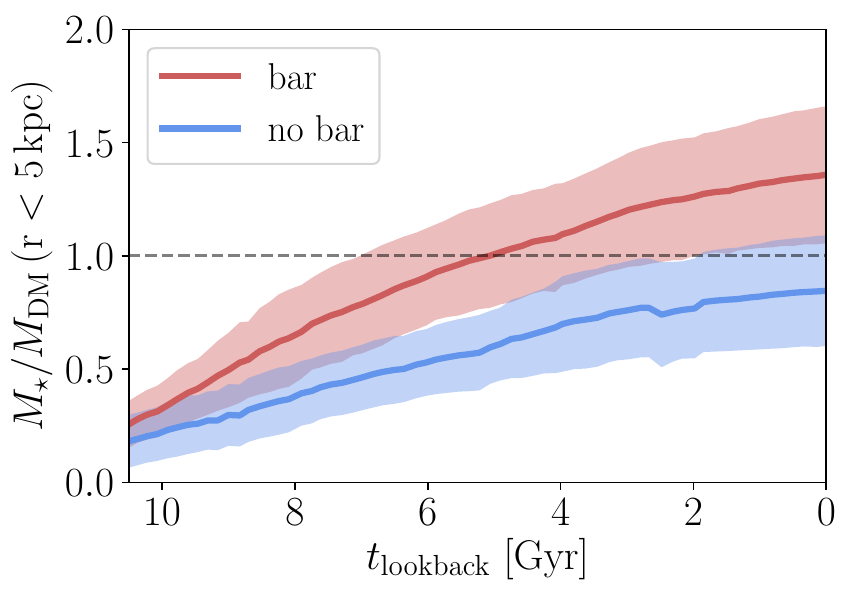}
\caption{The average (solid lines) and standard deviation (shaded regions) of the ratio of stellar-to-dark matter mass within 5\,kpc as a function of time for the barred galaxies (excluding the interacting galaxies), in red, and for the unbarred galaxies, in blue. On average, for barred galaxies, the stellar mass dominates over the dark matter mass within the central 5\,kpc by $z\sim 0$, which is in contrast to unbarred galaxies which tend to be more dark matter dominated over all cosmic times.} 
\label{fig:formtime}
\end{figure}

\begin{figure}
\centering
\includegraphics[width=0.45\textwidth]{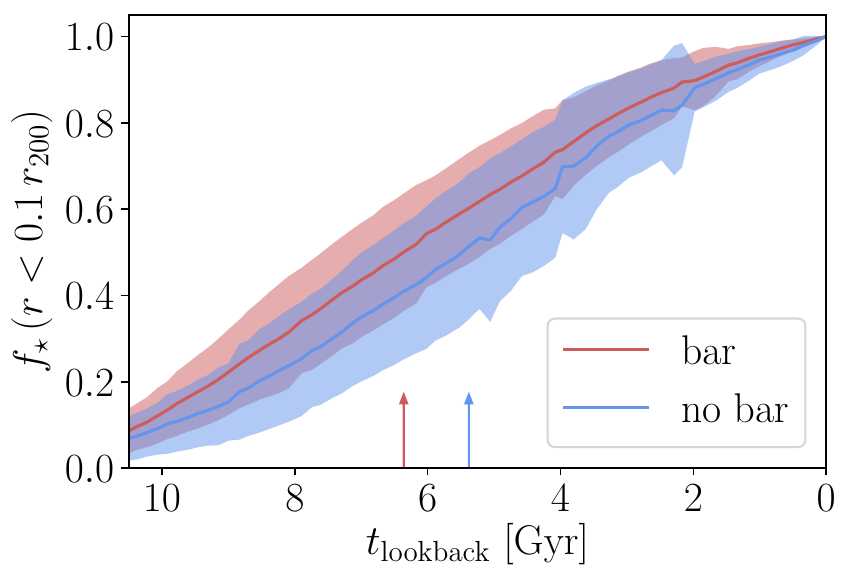}
\caption{The build-up of stellar mass in barred (without interacting) and unbarred galaxies as a function of time. The solid lines show the average values for the samples and the shaded regions the 1$\sigma$ dispersion. Barred galaxies tend to build up their stellar mass more rapidly than unbarred galaxies, with the time at which they build up 50\% of their stellar mass roughly 1\,Gyr earlier than for unbarred galaxies (indicated by the red and blue arrows respectively).} 
\label{fig:formtime50}
\end{figure}

\begin{figure}
\centering
\includegraphics[width=0.42\textwidth]{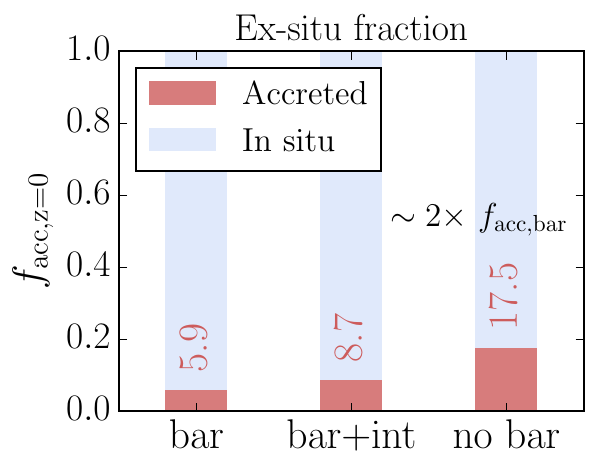}
\caption{Average fraction of accreted stars in the `bulge', at $z=0$, for the barred (without and with interacting galaxies, left and middle) and unbarred (right) galaxies in our sample. Unbarred galaxies have on average $\sim 2 \times$ more ex-situ stars in their bulge than the barred sample (without interacting galaxies) and 60\% more ex-situ stars in the bulge than the total barred sample.} 
\label{fig:facc}
\end{figure}

\begin{figure*}
\centering
\includegraphics[width=0.95\textwidth]{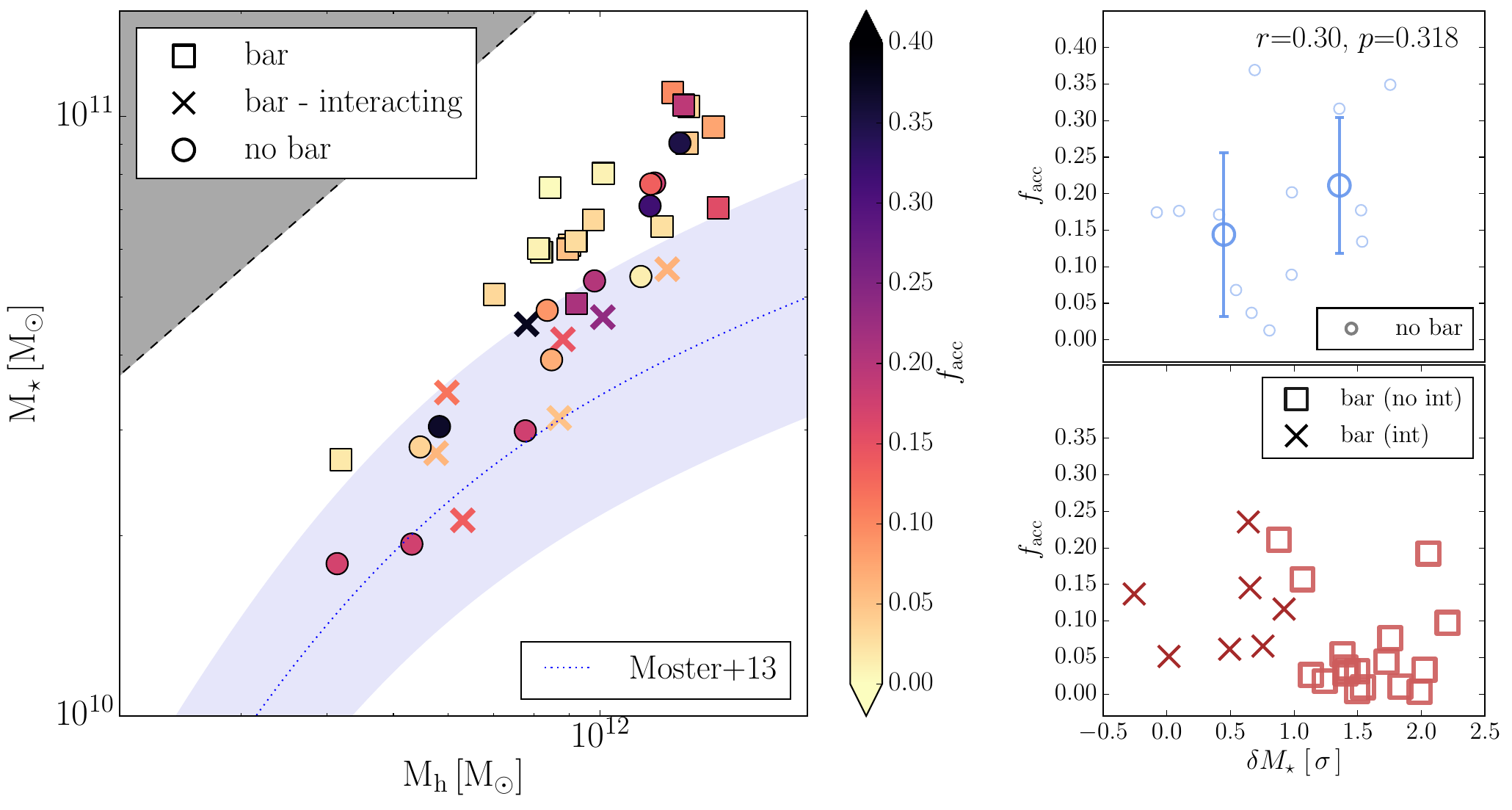}
\caption{The relation between the $M_{\star}-M_{\rm h}$ plane and the fraction of ex-situ stars in the central 4\,kpc. \emph{Left:} The $M_{\star}-M_{\rm h}$ plane colour-coded by the accreted fraction of stars in the bulge. \emph{Right:} The fraction of accreted stars in the bulge as a function of the offset from the abundance matching relation for the unbarred galaxies (top) and the barred galaxies (bottom). We see that galaxies that are more offset from the abundance matching relation tend to have higher ex-situ fractions, indicating that the ex-situ bulge contributes to suppressing bar formation in baryon-dominated systems.} 
\label{fig:amwfacc}
\end{figure*}

\begin{figure}
\centering
\includegraphics[width=0.49\textwidth]{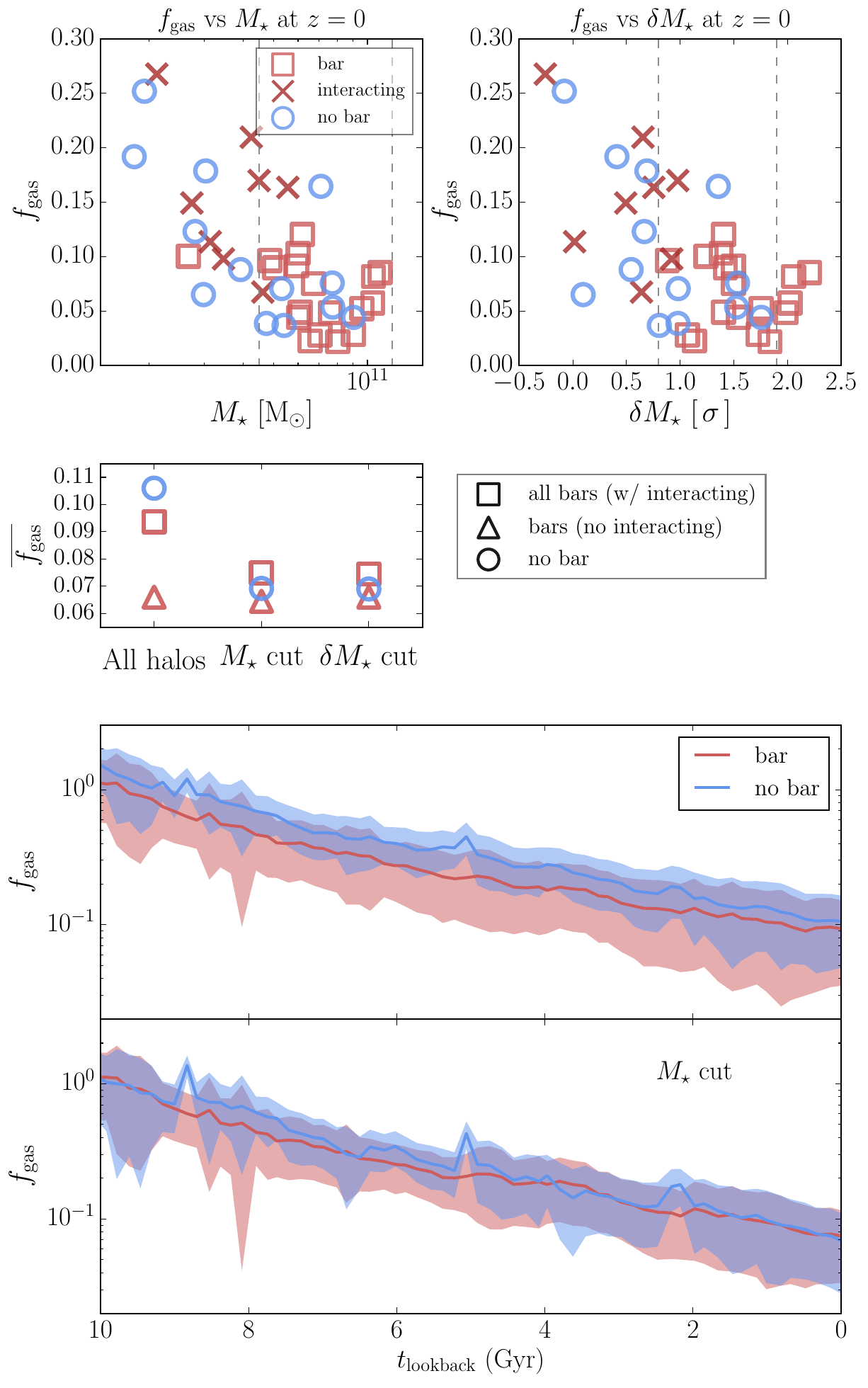}
\caption{The effect of gas fraction on bar formation: \emph{First row:} At $z=0$, the gas fraction for each halo as a function of stellar mass (left) and as a function of the baryon-dominance of the galaxy (right). The vertical dashed lines indicate the range in which galaxies are considered for the averages shown in the panel below. \emph{Second row:} The average gas fraction for various samples: all galaxies, galaxies with the $M_{\star}$ and $\delta M_{\star}$ cuts shown above, separating into all the bars, bars excluding interacting galaxies and unbarred galaxies.
\emph{Third row:} The gas fraction for all barred and unbarred galaxies in our sample across time. \emph{Fourth row:} As above, but for galaxies within the mass range shown in the top left panel. If we consider galaxies in a given mass bin, the gas fractions are the same for barred and unbarred galaxies.
} 
\label{fig:fgas_frac}
\end{figure}

\begin{figure}
\centering
\includegraphics[width=0.49\textwidth]{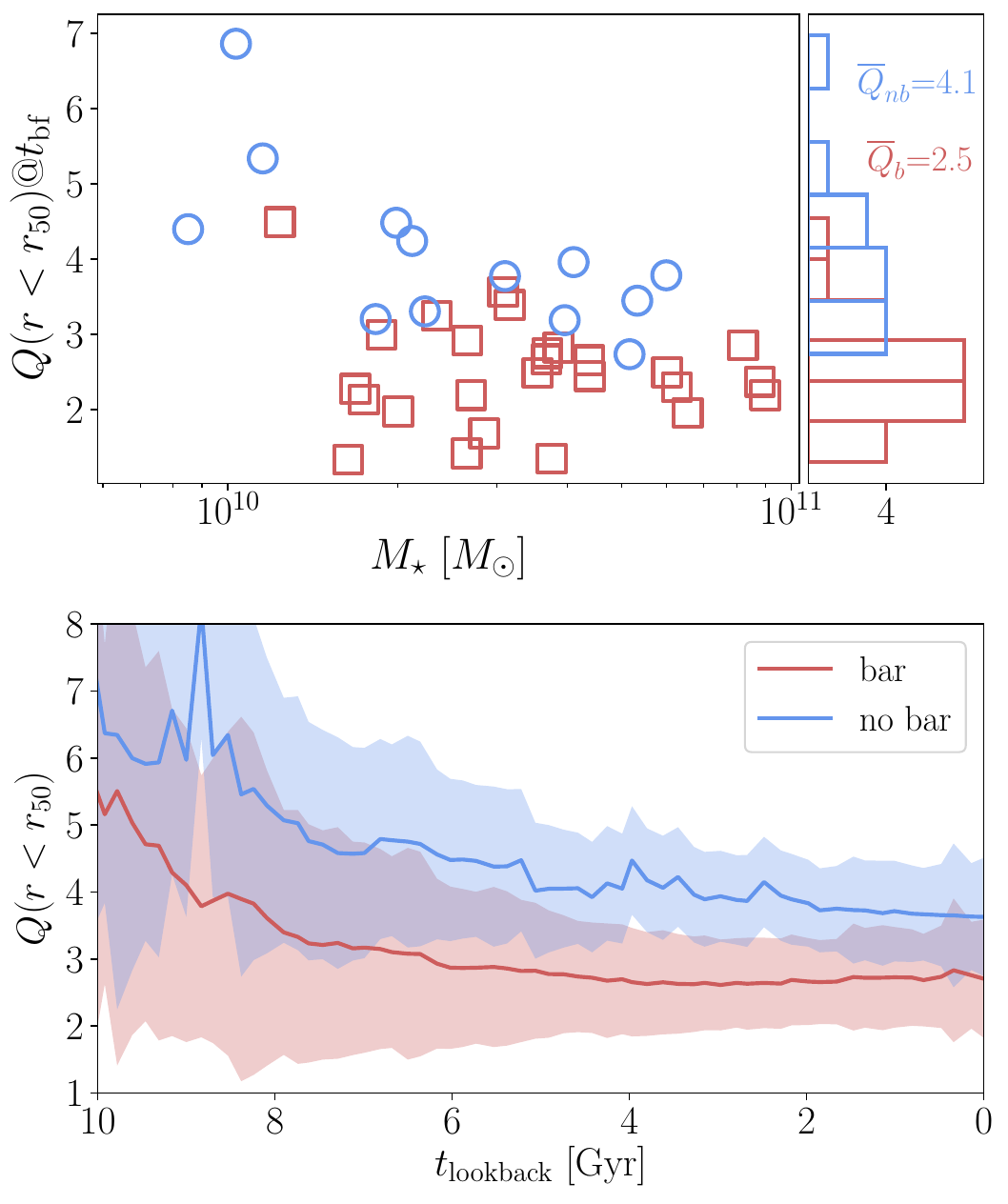}
\caption{
The Toomre Q parameter for the galaxies in our sample. \emph{Top:} Toomre Q inside stellar $r_{50}$ at the time of bar formation, for barred (red squares) and unbarred (blue circles) galaxies, as a function of the stellar mass of the galaxy at that time (note that for the unbarred galaxies we take these values at the time corresponding to the average bar formation time -- see text). Unbarred galaxies tend to have higher Toomre Q than barred galaxies at the time of bar formation for all stellar masses. \emph{Bottom:} The average Toomre Q inside $r_{50}$ for barred (red) and unbarred (blue) galaxies in our sample as a function of time.} 
\label{fig:toomreq}
\end{figure}

It is well established that the ratio of the stellar disc to the dark matter halo mass within a spiral galaxy is a determining factor in how unstable the galaxy will be to the formation of a bar (e.g. \citealt{OstrikerPeebles1973,Combesetal1990,Athanassoula2002,Fujiietal2018}).
Here we explore how the baryon-dominance of a galaxy affects the formation of a bar in the Auriga cosmological simulations. 

As discussed in Section \ref{sec:globalprops}, in the bottom panel of Fig.~\ref{fig:bars_vs_fbd} we compare the bar formation time to the time at which the galaxy has at least 80\% of the total mass within 5\,kpc dominated by the stellar component (which we define as the time at which the galaxy becomes baryon-dominated). We find that there is a correlation -- albeit with significant scatter -- between the time at which the bar forms and the time at which the galaxy becomes baryon-dominated, suggesting that once a disc becomes baryon-dominated it will tend to form a bar.

To further explore the role baryon-dominance plays on bar formation in the cosmological context, in Fig \ref{fig:amall}, we show the stellar mass against the halo mass for the galaxies in our sample, which are labeled according to whether they are barred, unbarred, or have a bar but are also interacting at $z\sim 0$. We also plot the abundance matching relation from \cite{Mosteretal2013} for comparison. We find that barred galaxies tend to be displaced from the commonly used abundance matching relation, as compared to unbarred galaxies, which lie closer to the relation.  We quantify this offset from the abundance matching relation in the top right panel of Fig.~\ref{fig:amall}: we show the vertical displacement of the galaxy from the abundance matching relation (i.e. the over- or under-abundances in stellar mass as compared to the abundance matching relation, for a given halo mass), normalised by the $1\sigma$ scatter of the abundance matching relation of \cite{Mosteretal2013}. The barred galaxy sample is shown in red, the unbarred sample in blue, and the barred sample including the interacting galaxies with the dashed brown line. We see that barred galaxies tend to be more than 2$\sigma$ offset from the abundance matching relation, compared to unbarred galaxies which lie closer to the abundance matching relation. The vertical arrows indicate the median value for each population. It is worth noting, that this 'offset' from the abundance matching relation correlates to the baryon-dominance $f_{\rm bd}$ of the galaxy, as can be seen in Fig.~\ref{fig:fbd_vs_AMoffset}. The bottom right panel of Fig.~\ref{fig:amall} shows the $V_{\rm \star}$/$V_{\rm tot}$ of the barred and unbarred galaxies in our sample, and in black for the Milky Way (as determined by \citealt{BovyRix2013}). We see that barred galaxies have higher $V_{\rm \star}$/$V_{\rm tot}$ than unbarred galaxies at all radii. However, this is not as high as that for the Milky Way, suggesting that Milky Way-like galaxies in cosmological simulations might still be less baryon-dominated than massive spiral galaxies in the local Universe. This fact, i.e. that cosmological simulations are not as baryon-dominated as suggested by dynamical models of massive spiral galaxies in the local Universe (e.g. \citealt{Kranzetal2003,Weiner2001,Fragkoudietal2017a}), is commonly found in the literature \citep{Lovelletal2018,Postietal2019,PostiFall2021,Fragkoudietal2021}. 

In Fig.~\ref{fig:formtime} we explore the ratio of stellar-to-dark matter within 5\,kpc ($f_{\rm bd}$) as a function of time; we find that, already at earlier lookback times, the barred galaxies are more baryon dominated than the unbarred sample which remains dark matter dominated even at low redshifts. This trend also holds if we explore the baryon dominance within the half-mass radii of the galaxies ($r_{\rm 50}$), as is shown in Fig.~\ref{fig:bardomr50}, with the differences between the two populations being even more pronounced (due to the fact that the half-mass radii of barred galaxies are smaller than for unbarred galaxies in our sample).

In Fig.~\ref{fig:formtime50} we show the growth of the stellar mass of the barred (red) and unbarred (blue) galaxies as a function of time. As discussed in Section \ref{sec:analysis}, we calculate the stellar mass within 10\% of the virial radius of the galaxy. We find that barred galaxies assemble their stellar mass more rapidly than unbarred galaxies (see also \citealt{RosasGuevaraetal2020}). We quantify this by exploring when the samples assemble 50\% of their stellar mass, and find that barred galaxies assemble their stellar mass on average $\sim$1\,Gyr earlier than unbarred galaxies. Barred galaxies therefore tend to grow their stellar component more rapidly than unbarred galaxies, forming an early massive stellar disc, which is conducive to bar formation.

\subsection{The ex-situ bulge}
\label{sec:exsitu}

We now explore how the fraction of ex-situ (accreted) stars in the central regions of galaxies affects the formation of bars. The fraction of accreted stars in the inner regions can be thought of as a proxy for a merger-built hot (classical) component, since these accreted stars will be brought in via mergers. For example, \cite{Gargiuloetal2022} showed that simulated galaxies whose bulge has a high S\'{e}rsic index will also tend to have a larger fraction of ex-situ. As hot spheroids (either stellar or dark matter) have been shown to delay the formation of bars (e.g. \citealt{OstrikerPeebles1973,Athanassoula2002}), one might expect such an ex-situ component to have an impact on bar formation in cosmological simulations. We use the fraction of ex-situ stars rather than a kinematic characterisation of the bulge because commonly-employed kinematic estimates (such as the circularity parameter \citealt{Abadietal2003}) are affected by the presence of a bar (which induces large non-circular motions in the orbits of stars in the central region). This makes it challenging to separate the contribution of the spheroid from the bar. In what follows we consider the accreted fraction within the inner 5\,kpc, but we note that our results are qualitatively similar when considering different radii.

We start by comparing in Fig.~\ref{fig:facc} the fraction of accreted stars in the bulge, for the  barred and unbarred galaxies in our sample (see also \citealt{Fragkoudietal2020}). We compare unbarred galaxies to barred galaxies (excluding interacting galaxies at $z=0$) and find that unbarred galaxies have $\times$2 higher fraction of accreted stars; this supports the previous theoretical results that have found that massive classical bulges can impede bar formation. The difference in the accreted bulge fraction between barred and unbarred galaxies becomes somewhat less prominent if we include the interacting galaxies in the barred sample (see middle column of Fig.~\ref{fig:facc}), however the unbarred galaxies still have $\sim50\%$ higher accreted fraction in the bulge as compared to the total barred sample. Similar results were found previously in \cite{Fragkoudietal2020} for the Auriga galaxies and in \cite{Gargiuloetal2022} for the \textsc{IllustrisTNG50} simulations, who found that galaxies with a low fraction of ex-situ stars in their bulge are more likely to host a long-lived bar. 

We further investigate whether the accreted fraction can prevent galaxies from forming bars by exploring how this relates to the baryon-dominance of galaxies. As discussed in the previous section, galaxies that are more baryon-dominated are more likely to form a bar. However, there are some galaxies in our sample which \emph{are} baryon-dominated (and offset from abundance matching relation -- see Fig.~\ref{fig:amall}), which have not formed a bar by $z=0$, indicating that something may be preventing them from forming a bar. In Fig.~\ref{fig:amwfacc} we explore how the offset from the abundance matching relation depends on the accreted fraction of stars for unbarred galaxies. In the left panel we plot the $M_{\star}-M_{\rm h}$, colour-coded by the faction of accreted stars. We see that galaxies that are baryon-dominated, i.e. are offset from the abundance matching relation, which do not host a bar, also tend to have a higher fraction of accreted stars. This is shown explicitly in the top right panel of Fig.~\ref{fig:amwfacc}, where we plot the fraction of accreted stars versus the offset from the abundance matching relation for the unbarred galaxies in our sample. While there isn't a strong correlation between the two parameters, we find that galaxies which are more offset from the abundance matching relation tend to have a higher fraction of accreted stars. 
This seems to suggest that having a higher fraction of accreted stars could suppress the formation of the bar, even when the galaxy is baryon-dominated. On the other hand, there is no obvious trend for the barred galaxies with accreted fraction (see bottom right panel of Fig.~\ref{fig:amwfacc}), and as already mentioned, these galaxies tend to have lower accreted fractions.

\subsection{Gas Fraction}
\label{sec:gasfrac}

In Fig.~\ref{fig:fgas_frac} we show the fraction of gas in barred and unbarred galaxies at $z=0$ within a radius of $r= 5$\,kpc, in order to explore the effect it has on bar formation in cosmological simulations. We focus on the gas fraction within the inner regions of the disc where the bar is, as that is where it will have the largest effect on the bar dynamics.
In the top left column, we show the gas fraction as a function of stellar mass. In the top right column we show the gas fraction as a function of the offset from the abundance matching relation (which as discussed previously, correlates with the baryon-dominance). We see that there is an overall decrease in the gas fraction for higher stellar mass galaxies, while a similar decrease in gas fraction can be seen as a function of baryon-dominance, with both barred and unbarred galaxies occupying similar regions in these planes. Interestingly, in the top panel, we see that the barred galaxies that have lower masses (and high gas fractions) are also the ones that appear to have had a recent interaction. 

In the middle left panel of the figure we show the average gas fraction in the barred sample, with and without interacting galaxies (square and triangle points respectively) and unbarred galaxies (circle), in the leftmost point (labelled `All halos'). We find that when considering all galaxies in our sample, unbarred galaxies tend to have higher gas fractions (by about $\sim13\%$) than barred galaxies. This is accentuated (to $\sim$60\% higher gas fraction) when comparing unbarred galaxies to barred galaxies which are not interacting. This might lead one to conclude that the gas fraction is higher for unbarred galaxies;
however, as there is a dependence of bar fraction on stellar mass -- see e.g. in Fig.~\ref{fig:summarysample} -- we repeat this measurement by factoring out the mass dependence. That is to say, since there is an anti-correlation between gas fraction and stellar mass (i.e. low mass galaxies tend to have higher gas fractions), it is unclear whether the average higher gas fractions of unbarred galaxies could be due simply to the fact that low mass galaxies in our sample tend to be unbarred. To factor out the stellar mass dependence, we apply a mass cut as indicated by the vertical dashed lines in the top left panel, and recalculate the average gas fraction for our three samples in that mass range; this is shown in the middle left panel (labelled `$M_{\rm \star}$ cut'). We also explore the average gas fraction for a given baryon-dominance cut (labelled `$\delta M_{\star}$ cut'), with the restricted range shown with the vertical lines in the top right panel. We find that both when restricting our sample to a smaller range in stellar mass or baryon-dominance, the average gas fractions of barred and unbarred galaxies are very similar, differing by less than 3\%.

In the two bottom panels of Fig.~\ref{fig:fgas_frac} we show the gas fraction within 5\,kpc for barred and unbarred galaxies as a function of time. In the second to last panel, we see that unbarred galaxies tend to have slightly higher gas fractions across time. However, in the bottom panel we see that, if we restrict the sample to galaxies within the same mass bin (as in the top left panel of Fig.~\ref{fig:fgas_frac}), we find similar gas fractions for both barred and unbarred galaxies across all cosmic times.

We therefore conclude that, gas fraction does not seem to be a driving factor of bar formation in and of itself, since if we restrict our samples to a given stellar mass, or indeed a given baryon-dominance, barred galaxies do not tend to be more gas rich. Rather the difference in average gas fractions in the barred and unbarred sample over all stellar masses, seems to be a consequence of the dependence of gas fraction on stellar mass, which sets how baryon-dominated the galaxies are.

\subsection{The Toomre Q parameter}
\label{sec:veldisp}

A parameter which is known to play an important role in the formation of bars is the in-plane radial velocity dispersion of the stellar component (e.g. \citealt{CombesSanders1981,AthanassoulaSellwood1986}), which affects how cold the disc is, and therefore its stability. This has often been explored in the literature by examining the dependence of bar properties on the Toomre $Q$ parameter \citep{Toomre1964},

\begin{equation}
    Q = \frac{\sigma_r \kappa}{3.36 G \Sigma_{\star}}
\end{equation}

\noindent where $\sigma_r$ denotes the radial velocity dispersion of the stellar disc, $\kappa$ the epicyclic frequency, $G$ Newton's gravitational constant and $\Sigma_{\star}$ is the surface density of the stellar disc.

In the top panel of Fig.~\ref{fig:toomreq} we show the average Toomre $Q$ parameter within the stellar half mass radius of the barred and unbarred galaxies in our sample at the time of bar formation, as a function of stellar mass (for the unbarred galaxies, we calculate this parameter at 4.6\,Gyr, which corresponds to the average time of bar formation for the galaxies in our sample). We see that the unbarred galaxies have a higher Toomre $Q$ parameter than the barred galaxies. This occurs for all stellar masses.

In the bottom panel of Fig.~\ref{fig:toomreq} we plot the average and 1$\sigma$ dispersion for $Q$ within the half-mass radius for barred (red) and unbarred (blue) galaxies, as a function of time. We see that unbarred galaxies tend to have higher $Q$ across all times, although the difference between the two samples decreases slightly at lower redshifts. 

We therefore find that the in-plane radial velocity dispersion -- and therefore the stability of the disc -- does play a role in determining whether a galaxy will be barred or unbarred in cosmological simulations. As the galaxy's formation history will determine the radial velocity dispersion within the galaxy (both in the innermost regions and in the disc), the relation between the galaxy's formation history and Toomre Q is worth exploring in future work.

\section{Discussion}
\label{sec:discussion}

\begin{figure*}
\centering
\includegraphics[width=0.4\textwidth]{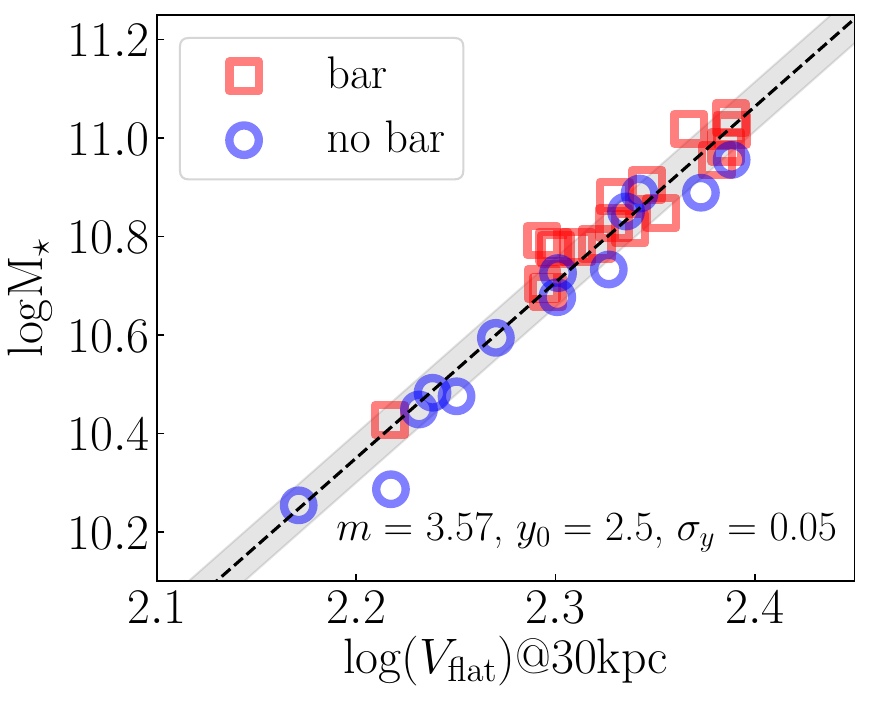}
\includegraphics[width=0.4\textwidth]{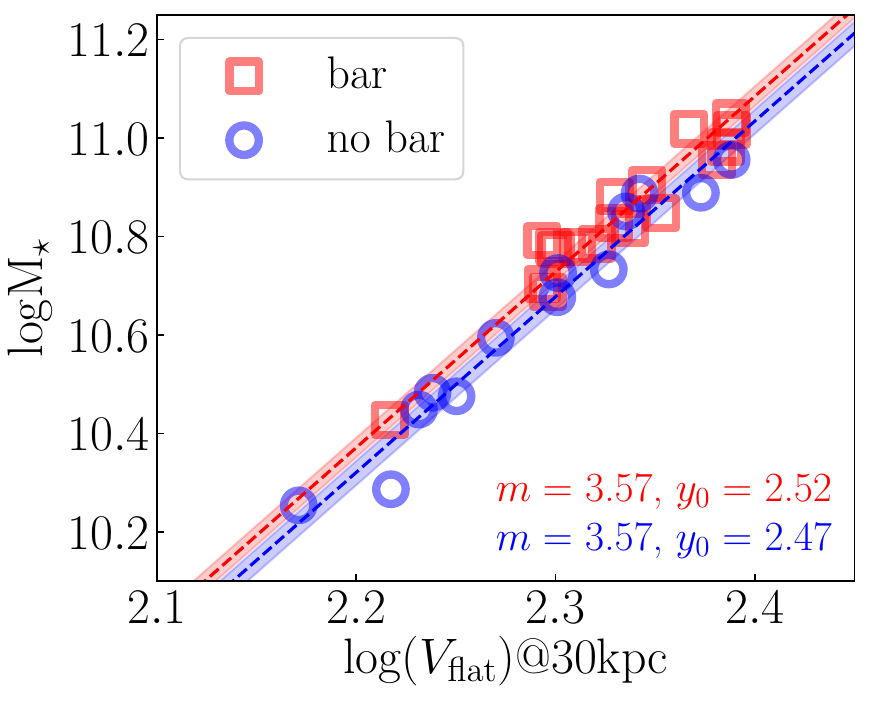}
\caption{The stellar Tully-Fisher relation at $z=0$ for galaxies in Auriga barred galaxies (excluding interacting) and unbarred galaxies. \emph{Left:} The grey dot-dashed line shows the relation given by fitting all galaxies together, while the shaded region indicates the intrinsic vertical scatter $\sigma_y$. \emph{Right:} The red (blue) dashed line shows the relation when fitting only the barred (unbarred) galaxies. The shaded regions indicate the uncertainties from boostrapping (see text).We find that the TF relation for barred galaxies is offset from that of the unbarred galaxies.} 
\label{fig:tfrelation}
\end{figure*}

\begin{figure*}
\centering
\includegraphics[width=0.99\textwidth]{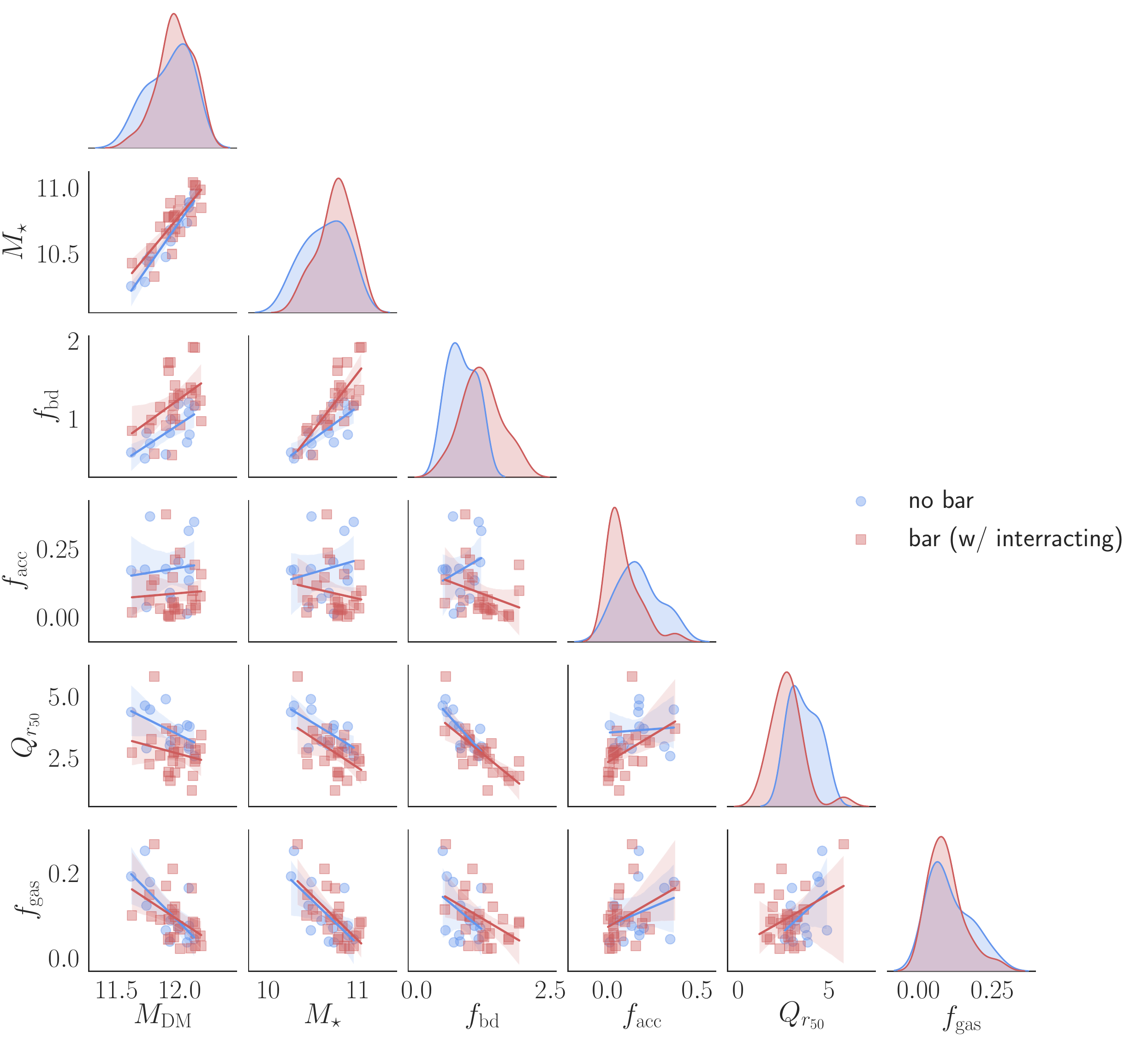}
\caption{The multi-dimensional parameter space of bar formation: dark matter halo mass ($M_{\rm DM}$), stellar mass ($M_{\star}$), baryon-dominance ($f_{\rm bd}$), fraction of accreted stars in the central bulge regions ($f_{\rm acc}$), the Toomre parameter within the half-mass radius ($Q_{r_{50}}$) and the gas fraction ($f_{\rm gas}$), all at $z=0$. The red (blue) points indicate the barred (unbarred) sample, including interacting galaxies. We see that the barred-unbarred populations separate out for the baryon dominance, accreted bulge fraction and Toomre $Q$ parameters.} 
\label{fig:cornerint}
\end{figure*}

In what follows we discuss the implicatios of our results in terms of the baryon-dominance of galaxies on the Tully-Fisher (TF) relation (Section \ref{sec:tfrel}) and what our findings about the evolution of bars imply about bars at high redshift (Section \ref{sec:predhighz}).

\subsection{Implications for the Tully-Fisher relation}
\label{sec:tfrel}

The Tully-Fisher relation (TF; \citealt{TullyFisher1977}) is an empirical relation between the rotation velocity of galaxies  -- often the radius at which the rotation curve flattens out -- versus their luminosity. This relation has low intrinsic scatter, making it a powerful tool for determining distances to galaxies, while it has also proved useful for testing galaxy formation models in $\Lambda$CDM \citep{MoWhite1998} and theories of modified gravity, such as Modified Newtonian Dynamics, or MOND (see e.g. \citealt{Milgrom1983,McGaugh2012}). 
In Fig.~\ref{fig:tfrelation} we explore the consequences of our findings in Section \ref{sec:baryondom} -- i.e. the fact that barred galaxies tend to be more baryon-dominated than unbarred galaxies -- on the stellar TF relation (STFR; in Appendix~\ref{sec:append} we show results for the Baryonic Tully Fisher relation). We obtain the circular velocity at $z=0$ of the galaxies in our sample at 30\,kpc -- which corresponds roughly to the flat part of the rotation curve for the galaxies we are exploring, i.e. $V_{\rm flat}$. We then fit the relation $\log{M_{\star}}=m \log{V_{\rm flat}} + y_0$ to our sample of barred galaxies (excluding interracting galaxies) and unbarred galaxies. In the left panel of Fig.~\ref{fig:tfrelation} we show a least squares fit to this sample (grey dashed curve). The vertical scatter of the relation in our sample is $\sigma_y=0.05$dex, as indicated with the grey shaded region. 
In the right panel of the Figure we fit separately the barred (red) and unbarred (blue) samples. For these fits, we fix the slope of the relation to be that obtained for the total sample and we perform bootstrapping to assess the uncertainties in the fits. The 95\% percentile uncertainty in the intercept, as obtained form our fits, is shown with the shaded regions.
We find an offset in the STFR of the barred galaxies as compared to that of the unbarred galaxies, of the order of $\Delta y_0 = 0.05$dex. This offset is comparable to the intrinsic scatter of the entire sample, suggesting that, while challenging, it could in principle be observed.

We now turn our attention to what observational studies have found regarding the barred and unbarred TF relation. There have been few of these carried out in the literature, often finding antithetical results. For example, the study of \cite{Sakaietal2000} found that the TFR of barred galaxies is offset from the unbarred one, suggesting that barred galaxies have maximal discs. The subsequent study of \cite{Courteauetal2003} explored this in two separate samples, the Shellflow and SCII surveys.  \cite{Courteauetal2003} found a difference in the TFR for barred and unbarred galaxies in the Shellflow survey, while they did not find any offset in the SCII sample. Due to the marginally larger number of galaxies in the latter sample, the authors concluded that there is no difference between the barred and unbarred TFR. We note that these studies did not have resolved HI rotation curves, which would therefore increase the scatter in the relation (e.g. \citealt{Ponomarevaetal2017}). Therefore, the question of whether the barred TFR is offset from the unbarred one in observational samples still remains, and would be worth exploring with more recent and larger samples. 

It is worth emphasizing, that should such a trend be confirmed in the observed barred and unbarred TFR, this would be a strong indication for the existence of dark matter, since such an offset cannot be naturally explained in a modified gravity framework, such as in Modified Newtonian Dynamics (MOND; \citealt{Milgrom1983,McGaugh2012}). In a MONDian framework, the $V_{\rm flat}$ would be exactly proportional to the enclosed mass, and as such the relation should not have any intrinsic scatter (see e.g. \citealt{McGaughetal2000}). Finding an offset between barred and unbarred galaxies would be indicative of intrinsic scatter in the relation, which is more naturally explained in a framework with dark matter, i.e. where barred galaxies are more baryon-dominated in the inner regions, with the stellar component of the disc dominating over the dark matter halo.

\subsection{Bars at high-$z$}
\label{sec:predhighz}

A number of observational studies have explored the fraction of barred galaxies between redshifts of $z=0-1$, with most studies finding a declining bar fraction as a function of redshift \citep{Abrahametal1999M,Shethetal2008,Melvinetal2014, Simmonsetal2014}, although some studies find constant bar fractions (specifically for strong bars) \citep{Jogeeetal2004}. Studies using data from the Hubble Space Telescope (HST) have found that by $z=1$, the bar fraction drops to $\sim 10\%$ (e.g. \citealt{Simmonsetal2014}). Recent studies using JWST data have been revealing barred galaxies at $z > 1$ (e.g. \citealt{Guoetal2023}), and shedding new light on the bar fraction at these redshifts. For example, a recent study by \cite{LeConteetal2024} -- covering the redshift range $z=1-3$ with data from the Cosmic Evolution Early Release Science Survey (CEERS; \citealt{Finkelsteinetal2023}) -- found that compared to previous HST results, the fraction of bars is higher by a factor of $\sim2$ when considering the higher sensitivity JWST data, finding a bar fraction of $20\%$ at $z\sim1-2$. At $z\sim2-3$ the bar fraction drops to $15\%$. As shown in Fig.~\ref{fig:barfrac}, these trends are reproduced in the Auriga simulations, with a decreasing fraction of barred galaxies as a function of redshift. It is important to note that our sample covers a small mass range at $z=0$, and therefore our results do not make predictions for the behaviour of lower mass galaxies. 

Cosmological simulations with different galaxy formation models do not all agree on the bar fractiono at low and high redshfits. For example, some studies find a dearth of bars at high redshifts, which is exacerbated at low redshifts (e.g. \citealt{Reddishetal2022}). 
Other studies of bar fractions using `big box' cosmological simulations (e.g. \citealt{Peschkenetal2019,RosasGuevaraetal2022,Gargiuloetal2022}) find a flat, or rising bar fraction from $z=0$ to $z\sim1$, in contrast to most observational studies of bar fractions which find a decreasing bar fraction. One possible explanation for this mismatch (as mentioned in the latter studies) is that observational studies are not able to find short bars at higher redshifts, due to resolution limits. However, another possible explanation for this mismatch could be that the galaxy formation models employed affect the bar fraction across redshift. Whether the bar fraction remains flat/rising or decreases towards higher redshifts will therefore be an important point to settle in upcoming observational studies with higher spatial resolution and sensitivity, for example by using data from facilities such as JWST.

In the mass range probed by our simulations, we find that bars that form at higher redshifts tend to form longer than their local Universe counterparts. This occurs because bars at high redshifts tend to be triggered by interactions and/or mergers, which might cause the bars to form more `saturated', i.e. longer. As shown in Fig.~\ref{fig:bardiscratio} this results in these galaxies having a higher ratio of bar length relative to the disc size, as compared to local galaxies.  Therefore -- for stellar masses in the range probed by our sample -- we might expect to find that observed high redshift bars (e.g. with JWST) are relatively long, and that the bar length-to-disc size ratio is larger than that observed in the local Universe.

\subsection{The multi-parameter space of bar formation}
\label{sec:multiparam}

In this work, we have explored various galaxy properties which are thought to affect bar formation, in order to gain a better understanding of the role these parameters play when taken into account within the global cosmological context. This can help pave the way towards establishing the multi-parameter space that sets whether a galaxy will form a bar or not within the cosmological setting, which could have a number of useful applications (e.g. for determining whether a galaxy will form a bar in semi-analytic cosmological simulations). 

We attempt to summarise our findings in Fig.~\ref{fig:cornerint}, where we show the distribution of barred and unbarred galaxies in different combinations of the parameter space we have explored in the previous sections. We find that, as discussed, the main parameters that separate out the barred-unbarred population are mainly the baryon-dominance $f_{\rm bd}$, the fraction of accreted stars within the inner region (bulge) of the galaxy $f_{\rm acc}$ and the Toomre $Q$ parameter of the inner galaxy $Q_{r_{50}}$, although we note that there is significant overlap between in which galaxy will be barred or unbarred in either one of these properties separately. However, for some of properties, for example the gas fraction $f_{\rm gas}$ or the dark matter halo mass $M_{\rm DM}$, the barred and unbarred samples are completely overlapping. In future, this multi-dimensional parameter space could be used to build a semi-empirical model for bar formation within the cosmological context.

\section{Summary \& Conclusions}
\label{sec:summary}

In this work we study the formation and evolution of barred galaxies in the cosmological context, using the Auriga suite of cosmological zoom-in simulations, focusing on galaxies around the mass of the Milky Way (i.e. with halo masses between $0.5 \times 10^{12}-2 \times 10^{12} \rm M_{\odot}$ at $z=0$). We start by exploring the properties of barred galaxies and how these evolve over cosmic history. We then focus our attention on the question `which galaxies form bars?' -- i.e. we examine the various galaxy properties which are thought to play a role in the formation of bars. This enables us to determine the importance of these various properties within the cosmological context, in order to begin piecing together the multi-dimensional parameter space that drives bar formation. We summarise our results below.

In terms of the global properties of barred galaxies and on the evolution of bars, we find that: 
\begin{itemize}
    \item In Auriga, the fraction of barred galaxies decreases up to $z\sim1$, and then remains $\sim20\%$ out to $z\sim3$, consistent with the observed bar fraction found from recent JWST studies (Fig.~\ref{fig:barfrac}). The fraction of barred galaxies is higher in high mass galaxies across the redshifts explored.
    \item Below $z\sim2$, bars in the Auriga simulations are robust, long-lived structures. They have a wide range of formation times, with the oldest bar in our sample forming at lookback time of $t_{\rm lb} = 10.37$\,Gyr and the youngest forming at $t_{\rm lb} = 0.34$\,Gyr (e.g. Fig.~\ref{fig:summarya2}).
    \item At $z=0$, older bars tend to also be stronger, although with significant scatter in the relation (Fig.~\ref{fig:bars_vs_fbd}).
    \item The bar formation time, $t_{\rm bf}$, correlates with: i) the time at which the galaxy becomes baryon-dominated, ii) the time at which the galaxy assembles 50\% of its stellar mass, as well as with iii) the baryon-dominance of the galaxy at $z=0$; it does not correlate with the host galaxy's stellar mass $M_{\rm \star}$ at $z=0$ (e.g. Fig.~\ref{fig:bars_vs_fbd} and Fig \ref{fig:barformgalform}).
    \item Bars that form at higher redshifts in our simulations tend to be longer than those that form in the low redshift Universe (Fig.~\ref{fig:barlength}). We find that this is due to the fact that bars at high redshifts are formed just after a significant merger, which leads them to form with a saturated length (Fig.~\ref{fig:barlength_vtime}). 
    \item Barred galaxies at high redshifts have larger bar-to-disc size ratios, due to the fact that bars forming in the early Universe are longer and have smaller size stellar discs (Fig.~\ref{fig:bardiscratio}).
\end{itemize}

When exploring the physical properties that are important for determining whether a galaxy will form a bar or not, we find that:

\begin{itemize}
    \item Baryon-dominance is confirmed to play a prominent role in determining whether a galaxy will form a bar or not, with barred galaxies being on average more baryon-dominated from high redshifts, down to $z=0$. This leads to barred galaxies being offset from commonly-used abundance matching relations (e.g. \citealt{Mosteretal2013}) in the $M_{\rm \star}-M_{\rm h}$ plane (Figs. \ref{fig:amall} \& \ref{fig:formtime}). 
    \item Galaxies that form bars assemble their stellar mass more rapidly (on average $\sim3$\,Gyr earlier) than unbarred galaxies (Fig.~\ref{fig:formtime}).
    \item Galaxies that are offset from the abundance matching relation -- i.e. which are baryon-dominated -- but which do not have a bar at $z=0$, tend to have a higher fraction of accreted material in their bulge (Fig.~\ref{fig:amwfacc}). This suggests that galaxies with a larger ex-situ bulge are more stable against bar formation. On average, barred galaxies have a lower ex-situ bulge than unbarred galaxies at $z=0$ (Fig.~\ref{fig:facc}).
    \item Unbarred galaxies tend to have a larger Toomre Q parameter, and therefore more stable discs, both at high redshifts (i.e. around the time of bar formation) and at $z=0$ (Fig.~\ref{fig:toomreq}).
    \item On average, unbarred galaxies have higher gas fractions than barred galaxies. However, when controlling for stellar mass, the difference between barred and unbarred galaxies in terms of gas fraction disappears (Fig.~\ref{fig:fgas_frac}). This suggests that differences in gas fraction are a secondary effect which is due to differences in the stellar mass (coupled to the fact that bars are more common in high mass galaxies in our sample). For a given stellar mass and/or baryon-dominance, barred and unbarred galaxies have similar gas fractions. We therefore do not find that gas fraction plays a major role in itself, in whether or not a galaxy will be barred.
    \item We summarise our results on the multi-dimensional parameter space that drives bar formation in Fig.~\ref{fig:cornerint}, where we see that the barred and unbarred samples separate out best in baryon dominance ($f_{\rm bd}$), accreted bulge fraction ($f_{\rm acc}$) and Toomre $Q$ parameter.
\end{itemize}

The fact that barred galaxies tend to be more baryon-dominated has important consequences for the loci of barred and unbarred galaxies in the Tully-Fisher (TF) relation (see Fig. ~\ref{fig:tfrelation}). Namely, barred galaxies are offset from unbarred galaxies in the TF relation (i.e. for the same $V_{\rm flat}$, barred galaxies have a larger $M_{\star}$ or $M_{\rm baryonic}$). Confirming this offset for barred galaxies in the observed TF relation would have important implications for the existence of dark matter, since such an offset cannot be readily explained in a modified gravity (e.g. MOND) framework.

\section*{Acknowledgements}
The authors thank Marie Martig, Simon White, Federico Lelli and Kyle Oman for useful discussions, and David Campbell and Adrian Jenkins for generating the initial conditions and selecting the sample of the Auriga galaxies. 
FF is supported by a UKRI Future Leaders Fellowship (grant no. MR/X033740/1). RG acknowledges support from an STFC Ernest Rutherford Fellowship (ST/W003643/1). FAG acknowledges support from ANID FONDECYT Regular 1211370, by the ANID BASAL project FB210003 and acknowledges funding from the Max Planck Society through a “Partner Group” grant.
FM acknowledges support from the European Union - NextGenerationEU, in the framework of the HPC project – “National Centre for HPC, Big Data and Quantum Computing” (PNRR - M4C2 - I1.4 - CN00000013 – CUP J33C22001170001). This work was supported by STFC
[ST/X001075/1]. Part of the simulations of this paper used: the
SuperMUC system at the Leibniz Computing Centre (www.lrz.de), Garching,
under the project PR85JE of the Gauss Centre for Supercomputing e.V. (www.gauss-centre.eu); the Freya computer cluster at the Max Planck Institute for Astrophysics; and the DiRAC@Durham facility managed by the Institute for Computational Cosmology on behalf of the STFC DiRAC HPC Facility (www.dirac.ac.uk). The equipment was funded by BEIS capital funding via STFC capital grants ST/K00042X/1, ST/P002293/1, ST/R002371/1 and ST/S002502/1, Durham University and STFC operations grant ST/R000832/1. DiRAC is part of the National e-Infrastructure.

\section*{Data Availability}

All the data used, as well as parts of the auxiliary data produced for this paper, are publicly available to download at \url{https://wwwmpa.mpa-garching.mpg.de/auriga/data.html}

\bibliographystyle{mnras}
\bibliography{References}

\begin{appendix}
\label{sec:append}

\section{The evolution of unbarred galaxies}

\begin{figure}
\centering
\includegraphics[width=0.49\textwidth]{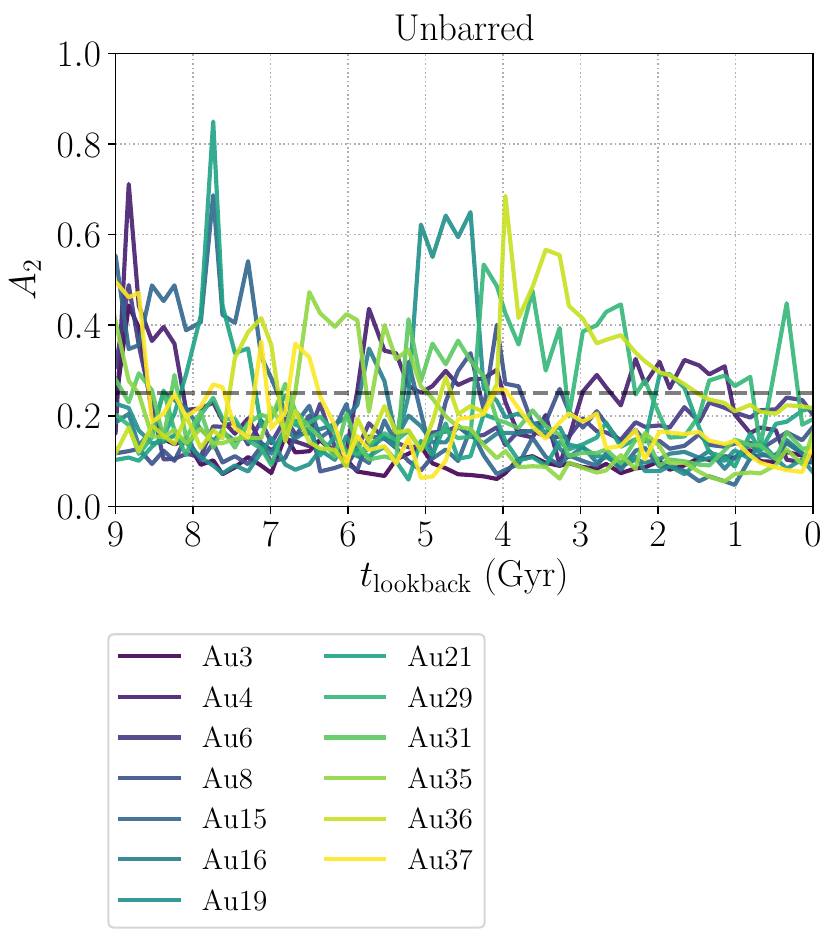}
\caption{The $m=2$ Fourier mode of the surface density as a function of lookback time for galaxies which are not classified as barred at $z=0$.} 
\label{fig:a2nobar}
\end{figure}

In Fig.~\ref{fig:a2nobar} we show the evolution of the bar strength $A_2$ as a function of redshift for the galaxies which do not contain a bar at $z=0$ in our sample, in order to explore if any of these galaxies have at some point hosted a bar. We examine in detail the galaxies in this sample that have peaks or spikes in $A_2$, to explore whether these are indeed due to the presence of a bar. We find that in all but one of these galaxies, the peaks in $A_2$ are not due to the presence of a stellar bar, but are transient effects due to interactions and/or mergers. The exception to this rule is Au36 which does indeed form a bar at a $t_{\rm lb}\sim 4$\,Gyr, which is subsequently gradually weakened by an interaction. This leads to the dissolution of the bar in this galaxy, which appears as an unbarred (or very weakly barred) at $z=0$ according to our definition (which requires $A_2>0.25$). We therefore see that in the Auriga simulations bars are long-lived structures, with only one case of a bar that is severely weakend after its formation.

\section{Relation between different estimates of baryon-dominance}

\begin{figure}
\centering
\includegraphics[width=0.45\textwidth]{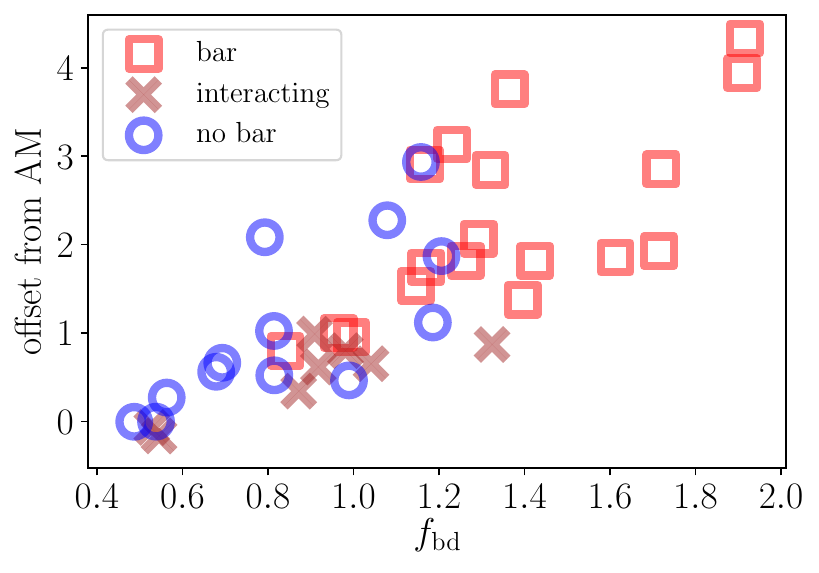}
\caption{The relation between the ratio of stars to dark matter within 5\,kpc ($f_{\rm bd}$) versus the offset from the abundance matching relation ($\delta M_{\rm \star}$). The two quantities are correlated.} 
\label{fig:fbd_vs_AMoffset}
\end{figure}

\begin{figure*}
\centering
\includegraphics[width=0.95\textwidth]{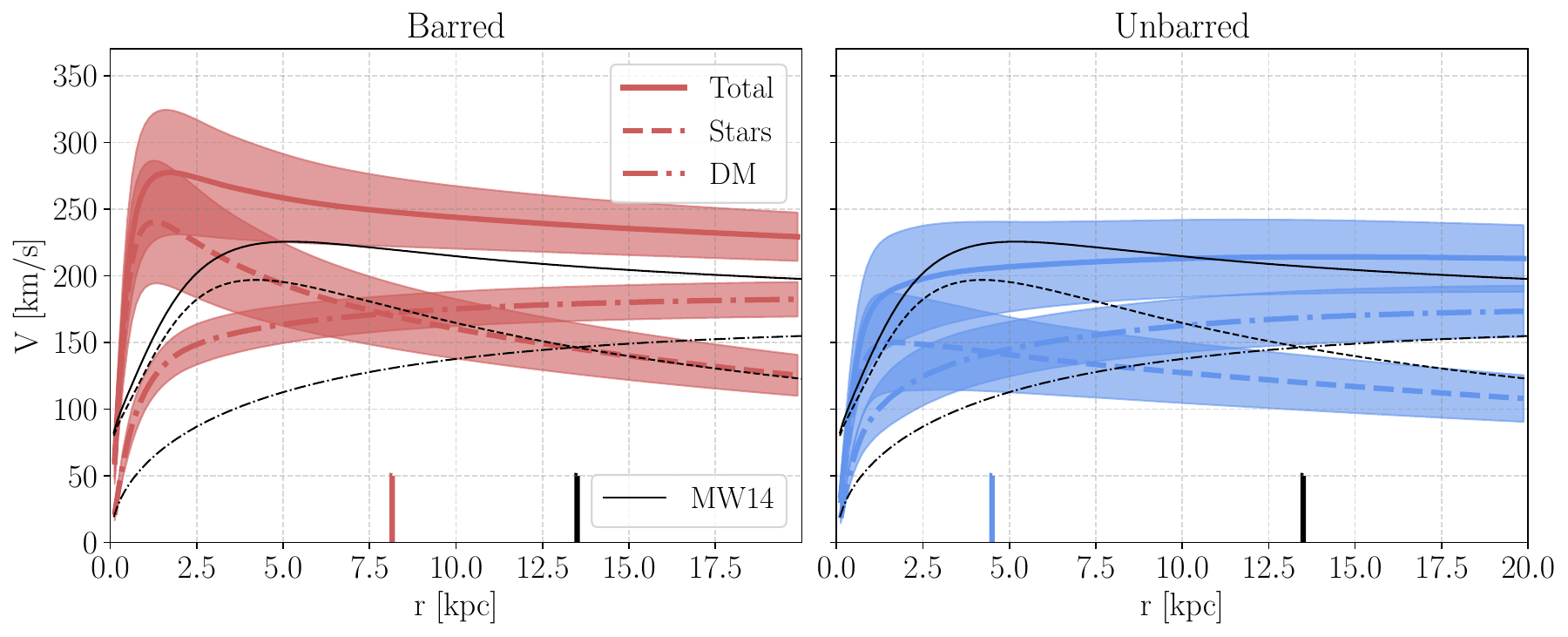}
\caption{Rotation curve of barred (left) and unbarred (right) galaxies in the Auriga simulation. We show the mean and $1\sigma$ dispersion for the total (solid), stars (dashed) and dark matter (dot-dashed). The contribution to circular velocity from the gas is not shown here but is included in $V_{\rm{tot}}$. The MW14 model is shown in black. The radius at which the dark matter begins to dominate over the stars is shown as vertical arrows on the bottom $x$-axis for the Auriga galaxies (red and blue) and the MW14 model (black).} 
\label{fig:rotbarnobar}
\end{figure*}

\begin{figure}
\centering
\includegraphics[width=0.45\textwidth]{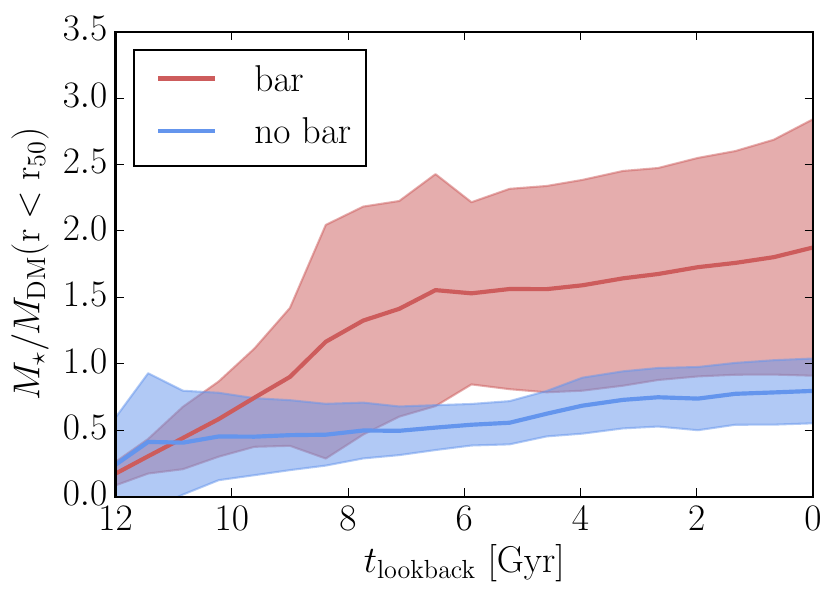}
\caption{The average (solid lines) and standard deviation (shaded regions) of the ratio of stellar-to-dark matter mass within the half-mass radius as a function of time for the barred galaxies (excluding the interacting galaxies), in red, and for the unbarred galaxies, in blue. On average, for barred galaxies, the stellar mass dominates over the dark matter mass within the half-mass radius by $z\sim 0$, in contrast to unbarred galaxies which tend to be more dark matter dominated over all cosmic times.} 
\label{fig:bardomr50}
\end{figure}

In Fig.~\ref{fig:fbd_vs_AMoffset} we show the relation between the ratio of stellar to dark matter within 5\,kpc ($f_{\rm bd}$) versus the offset from the abundance matching relation of \cite{Mosteretal2013} as defined in Section \ref{sec:baryondom} by $\delta M_{\rm \star,AM}$. We see that the two are correlated. The baryon-dominance $f_{\rm bd}$ of the barred galaxies can also be seen in Fig.~\ref{fig:rotbarnobar}, where the circular velocity curves for the barred sample (left) and the unbarred sample (right) are plotted. Solid lines give the average values for the two samples and the shaded regions the 1$\sigma$ dispersion. The thin black lines in both panels correspond to the MW14 model of \cite{BovyRix2013}. We see that the barred population is more baryon-dominated, i.e. the stellar component dominates in the inner regions, and they are overall more massive (higher circular velocity). The vertical lines give the point at which the stellar and dark matter components cross each other, which for the MW14 model occurs around $\sim13.5$\,kpc. For the Auriga barred population it occurs around 8\,kpc, while for the unbarred population it occurs at smaller radii (as expected for more dark matter dominated systems) at around 4\,kpc. We see that even though barred galaxies in Auriga are more baryon-dominated than unbarred ones, they are not as baryon-dominated as dynamical models for the Milky Way predict.

In Fig.~\ref{fig:bardomr50}, we show the baryon-dominance within the half-mass radius ($r_{\rm 50}$) as a function of time, for the barred and unbarred populations (solid line shows the average and the shaded region shows the 1$\sigma$ dispersion). This shows a similar trend to that discussed in Fig.~\ref{fig:formtime}, i.e. we find that barred galaxies are more baryon-dominated at earlier times than unbarred galaxies. Unbarred galaxies on the other hand stay dark matter dominated within the inner regions even down to $z=0$, within the half-mass radius.

\section{The Baryonic Tully Fisher Relation}
\begin{figure*}
\centering
\includegraphics[width=0.4\textwidth]{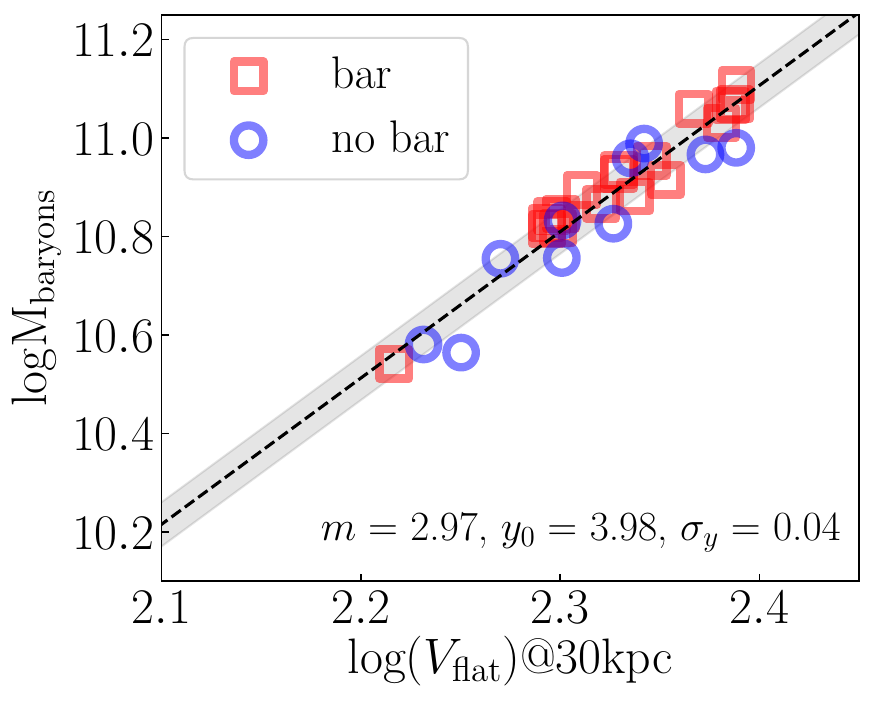}
\includegraphics[width=0.4\textwidth]{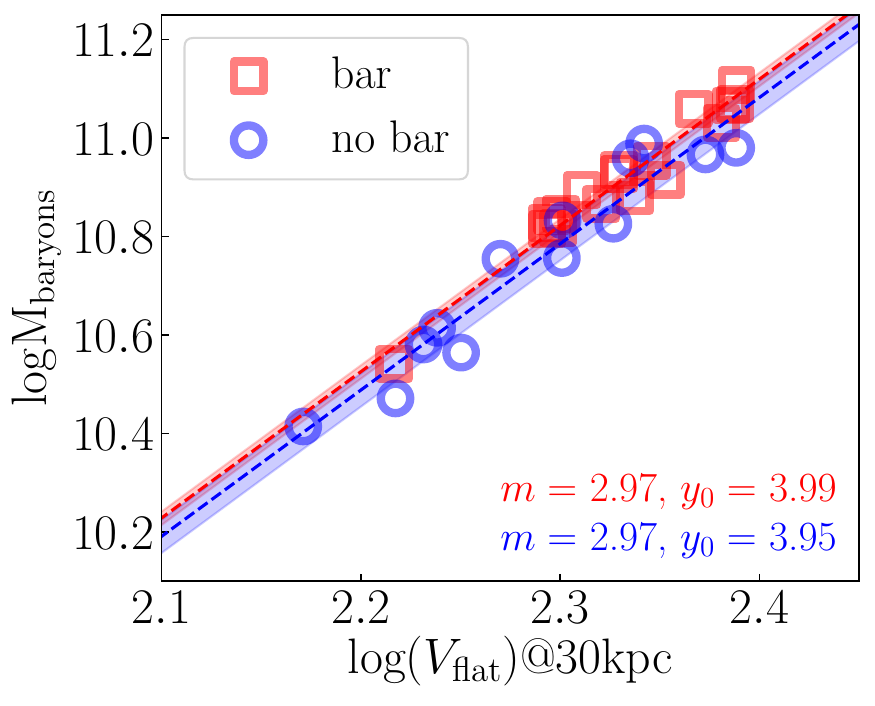}
\caption{The Baryonic Tully Fisher relation for galaxies at $z=0$ in Auriga. \emph{Left:} The grey dot-dashed line shows the relation given by fitting all galaxies together, while the shaded region indicates the intrinsic scatter $\sigma_y$. \emph{Right:} The red (blue) dashed line shows the relation when fitting only the barred (unbarred) galaxies. The shaded regions indicate the uncertainties from boostrapping (see text). The scatter is slightly reduced for the BTFR as compared to the STFR and the barred and unbarred BTFR relations are less separated.} 
\label{fig:btfrelation}
\end{figure*}
Here we extend our discussion in Section \ref{sec:tfrel} to the baryonic Tully Fisher Relation. Studies have shown that using the total baryonic mass of galaxies, i.e. combining both the gaseous and stellar component, results in even less intrinsic scatter in the relation, which is referred to as the Baryonic Tully Fisher relation (BTFR; e.g. \citealt{McGaughetal2000}). 
The BTFR has less than 0.05dex in orthogonal intrinsic scatter and $\sim0.075$dex in vertical scatter \citep{Lellietal2019}. 

In the left panel of Fig.~\ref{fig:btfrelation} we explore the BTFR for barred (excluding interacting) and unbarred galaxies in our sample. To obtain the baryon mass we add the gas mass of the galaxy -- which we obtain by summing all the gas within 0.1$r_{\rm 200}$ and $z<$2\,kpc -- to the stellar mass. In this mass regime the gas fractions are quite low (below 10\% for most galaxies) and as such, the results are similar to those found for the STFR, i.e. the barred galaxies are on average slightly offset from the unbarred galaxies, although the offset is less pronounced. We also find that the slope is decreased for the BTFR, i.e. we find a slope of $\sim3$, as opposed to the STFR where we found a slope of $\sim3.6$ (see \ref{sec:tfrel}). We note that it is not uncommon to find shallower slopes for samples that focus only on the high mass regime (see for example \citealt{Omanetal2016}).

\end{appendix}

\bsp	
\label{lastpage}
\end{document}